**Rate constants and product yields for the C + CH$_3$CHO reaction at low temperatures**

Kevin M. Hickson,[1,*] Jean-Christophe Loison,[1] and Valentine Wakelam[2]

[1] Univ. Bordeaux, CNRS, Bordeaux INP, ISM, UMR 5255, F-33400 Talence, France

[2] Univ. Bordeaux, CNRS, LAB, UMR 5804, F-33600 Pessac, France

**Abstract**

Reactions involving atomic carbon in its ground electronic state, C($^3$P), play an important role in astrochemistry due to high C-atom abundance levels. Here we performed a kinetic investigation of the reaction between C($^3$P) and acetaldehyde, CH$_3$CHO, determining rate constants for this process over the 50-296 K range. Measurements of the formation of atomic hydrogen, H($^2$S), were also performed to provide insight into product formation. Experiments were conducted using a supersonic flow reactor coupled with pulsed laser photolysis for C-atom generation and pulsed laser induced fluorescence in the vacuum ultraviolet range for the detection of both C($^3$P) and H($^2$S) atoms. Quantum chemical calculations of the ground triplet state potential energy surface of C$_3$H$_4$O were also performed to provide theoretical support for the measurements. The rate constants were large and temperature independent with an average value of $4.0 \times 10^{-10}$ cm$^3$ s$^{-1}$. This result is consistent with the theoretical results which predict either very low barriers or none at all on the underlying potential energy surface. Although experimental difficulties prevented the quantitative determination of H-atom formation, qualitatively, H-atom yields were very low with CH$_3$CH/C$_2$H$_4$ + CO as the major products based on the calculations. The influence of this reaction on interstellar chemistry was tested using a gas-grain model of dense interstellar clouds. These simulations predict that the C($^3$P) + CH$_3$CHO reaction decreases gas-phase CH$_3$CHO abundances by more than an order of magnitude at early and intermediate cloud ages, with a lower influence at typical dense cloud ages.

## 1 Introduction

As the fourth most common element, carbon and its compounds are widespread throughout the Universe. In the diffuse interstellar medium, where high photon fluxes are present, carbon is maintained in its cationic $C^+$ form due to its low ionization potential. As the density of these regions increases through gravitational collapse, the diminishing influence of external photons leads to the formation of neutral atomic carbon in its ground electronic $C(^3P)$ state (C hereafter). C-atoms have been observed in dense clouds such as TMC-1[1] and OMC-1[2] through their fine structure transitions at 492.162 and 809.345 GHz, clearly indicating their presence within these objects at high abundance levels, alongside major reservoir species of C such as CO. Astrochemical models simulating the evolution of the interstellar medium from a diffuse to a dense environment typically predict C-atom abundances about four orders of magnitude less than $H_2$ in the early and intermediate stages of cloud evolution.

Considering the large relative abundance of C in dense molecular clouds, its chemical pathways could play a crucial role in the formation and destruction of interstellar complex organic molecules (iCOMS) in particular. Indeed, C-atoms have been shown to react rapidly with unsaturated hydrocarbons such as ethene, $C_2H_4$, over a wide range of temperatures[3,4] to form larger complex organic species through addition-elimination type processes such as[5,6]

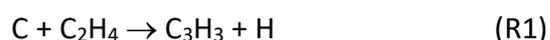
$$C + C_2H_4 \rightarrow C_3H_3 + H \quad (R1)$$

In contrast however, the products derived from C-atom reactions with other iCOMs, such as those containing heteroatoms (methanol, $CH_3OH$,[7] and acetonitrile, $CH_3CN$,[8] for example) do not always lead to an increase in the carbon chain length, with C-atoms reacting with $CH_3OH$ to form fragments (notably $CH_3$ + HCO), while C reacts with $CH_3CN$ by C atom addition followed by H atom elimination. In this respect, it is important to examine the reactions of atomic carbon with a range of iCOMs, to establish reactivity trends for specific types of reactions. In this respect, over the last few years we have performed a series of studies of C-atom reactions[7-15] with a range of O- and N-bearing molecules, combining experimental kinetic studies (measurements of the rate constants and product yields as a function of temperature) with theoretical work in many cases (calculations of the intermediates, transition states and possible product channels) to provide a detailed description of the individual reactive processes.

Of all the various iCOMs to have been identified so far, acetaldehyde, $CH_3CHO$ is one of the most widely observed. $CH_3CHO$ was first detected in Sgr B2 and Sgr A in 1973 through its emission line at 1065 MHz.[16] It has since been observed in numerous interstellar objects, including hot corinos, [17] starless cores[18] and in the prototypical dark cloud TMC-1 with an abundance of $2.7 \times 10^{-10}$ relative to $H_2$; a value taken from the statistical analysis of Gratier et al. [19] based on the TMC-1 spectral survey of Kaifu et al. [20] A more recent survey of 31 starless and prestellar cores in the Taurus Molecular Cloud derived $CH_3CHO$ abundances relative to $H_2$ within the range $(0.6-3.9) \times 10^{-10}$. [21] In terms of its gas-phase formation routes, astrochemical databases such as the Kinetic Database for Astrochemistry (KIDA)[22] indicate that $CH_3CHO$ can be formed by several neutral-neutral reactions such as those between atomic oxygen and the $C_2H_5$ and $C_3H_7$ radicals and also through the dissociative electron recombination reactions of several $C_2$ and $C_3$ bearing cations such as $CH_3CHOH^+$, $C_2H_5OH^+$, $C_3H_6OH^+$, etc. The O + $C_2H_5$ reaction is currently the most important gas-phase source of $CH_3CHO$ according to astrochemical models based on the kida.uva.2024 network. Gas-phase reactions however are not considered to be major sources of interstellar $CH_3CHO$, except maybe in dense molecular clouds. Instead, $CH_3CHO$ is thought to be mostly synthesized on interstellar ices before injection into the gas-phase although the exact formation mechanisms and desorption mechanisms are not well known. Although the radical recombination reaction between $CH_3$ and HCO is a potentially important pathway for $CH_3CHO$ formation on interstellar ices, theoretical studies[23] have shown that this process is disadvantaged by the geometrical constraints imposed by the water Ice versus and the most favored exit channel is exchange of H-atoms by the two radicals leading to the formation of alternative products $CH_4$ and CO. It has been suggested from theoretical studies that a potentially important formation route could be the condensation of atomic carbon on CO rich ices, leading to the formation of CCO. [24, 25] Hydrogenation reactions of the CCO radical could then lead to the formation of $CH_3CHO$ in addition to other molecules with a C-C-O backbone. At the present time, the observed $CH_3CHO$ abundances in dense interstellar regions are reasonably well reproduced by astrochemical models[26] despite the large uncertainty surrounding the mechanisms leading to its formation.

With regard to the possible reaction between atomic carbon and acetaldehyde, Husain and Ioannou[27] studied the kinetics of the gas-phase reactions between C-atoms and several aldehydes and ketones including $CH_3CHO$ at room temperature using a flash photolysis apparatus in the vacuum ultraviolet wavelength range, coupled with resonance absorption spectroscopy to detect atomic carbon. They measured a very large rate constant of $(5.4 \pm 0.3) \times 10^{-10}$ $cm^3$ $s^{-1}$ for this reaction. Although no previous theoretical studies of the C + $CH_3CHO$ reaction could be found in the literature, atomic carbon is predicted to undergo a barrierless reaction with formaldehyde, $H_2CO$, forming mostly H + HCCO and $CH_2$ + CO as products.[28, 29]

Here, we present the results of an experimental and theoretical investigation of the reaction between C-atoms and $CH_3CHO$ to improve our understanding of the interstellar chemistry of acetaldehyde and related molecules. Experimentally, a supersonic flow reactor was employed to conduct this investigation. The apparatus was coupled with a pulsed photolysis laser for C-atom generation and pulsed laser induced fluorescence (LIF) for detection purposes, allowing us to investigate the kinetics of the C + $CH_3CHO$ reaction over the 50-296 K temperature range. In addition, measurements of the H-atom yields of this reaction were also performed to improve our understanding of the major product channels. Theoretically, calculations of the possible products, intermediates and transition states over the triplet potential energy surface of the $C_3H_4O$ system were also performed, allowing us to provide more accurate information regarding the most probable pathways of the C + $CH_3CHO$ reaction. Finally, the rate constants derived in the present work were included in a gas-grain astrochemical model, updated for a better description of the chemistry of $CH_3CHO$ and related species.

The paper is organized as follows. The experimental and theoretical methods are described in sections 2 and 3 respectively, while the results of this work are presented in section 4. The astrochemical simulations and the implications of this study for interstellar dimethyl ether and related species are given in section 5, followed by our conclusions in section 6.

**2 Experimental Methods**

The experiments described in this work were carried out using a supersonic flow reactor, also known by the acronym CRESU (Cinétique de Réaction en Ecoulement Supersonique Uniforme). The CRESU technique provides a suitable means for avoiding the condensation of gaseous species at low temperature through the use of Laval type nozzles. These axisymmetric nozzles incorporate carefully designed upstream convergent and downstream divergent parts to produce a downstream gas flow with uniform (low) temperature, density and velocity for tens of centimetres from the nozzle exit. Due to the relatively high gas densities ($10^{16} - 10^{17}$ cm$^{-3}$) of the cold downstream gas, thermal equilibrium is maintained in the supersonic flow but condensation does not occur in these regions, unlike cryogenically cooled apparatuses due to the absence of contact of the cold gas flow with the walls of the reaction vessel when using the CRESU technique. The method was first employed to investigate ion-molecule reactions[30] before being adapted to the study of reactions between two neutral species. [31] A detailed description of the main features of the Bordeaux CRESU apparatus can be found in some of the earliest publications using this apparatus. [32, 33] Several modifications have since been integrated into the original apparatus, notably allowing us to study the kinetics of reactions involving atomic radicals which possess electronic transitions that typically fall in the vacuum ultraviolet (VUV) wavelength range. [7, 34-37] The current investigation employed three Laval nozzles based on argon and molecular nitrogen as the carrier gases, allowing four different flow temperatures (50 K, 75 K, 127 K and 177K) to be attained (one nozzle was used with both gases). The nozzle flow characteristics in terms of the flow temperatures, densities mass-flow rates and other details can be found in Table 1 of Hickson and Loison.[15] Room temperature experiments (296 K) were performed by using the reactor as a conventional slow-flow system in the absence of a Laval nozzle.

CH$_3$CHO was introduced into the flow through a bubbler system upstream of the Laval nozzle reservoir. In a similar manner to the system for introducing CBr$_4$ into the reactor (see below), a controlled flow of carrier gas (< 15 sccm) was first diverted into the CH$_3$CHO bubbler at room temperature. Then, the gas flow carrying acetaldehyde vapour was passed into a cold trap to ensure the carrier gas was saturated at the cold trap temperature (275 K). [38] The gas pressure in the bubbler and cold trap (500 $\leq$ P / Torr $\leq$ 600) were carefully regulated using a needle valve placed at the exit of the cold trap. The exit of the cold trap was connected to the reactor through a tube heated to 353 K to prevent any condensation between the cold trap and the reservoir. Once the CH$_3$CHO vapour reached the reservoir, it was diluted by the main carrier gas flow, so we assume no further condensation occurred from this point onwards.

C-atoms were created in-situ, within the supersonic flow by the pulsed laser photolysis of tetrabromomethane, CBr$_4$. Two different photolysis wavelengths were used in the present work. A large majority of the experiments following C-atom kinetics were performed using the output from a 10 Hz frequency quadrupled Nd:YAG laser at 266 nm as the source of atomic carbon. This beam was steered towards the reactor and passed through an afocal telescope to reduce the beam diameter from 10 mm to 5mm prior to entering the reactor through a quartz window at the Brewster angle to minimize window fluorescence. The beam propagated along the length of the supersonic flow and through the Laval nozzle throat before exiting the reactor through a second quartz window at the Brewster angle attached at the back of the reservoir. The exiting beam was captured by a laser energy meter allowing continuous monitoring of the photolysis pulse energy which was fixed in the range 30-35 mJ. A few experiments following C-atom kinetics and the majority of H-atom detection experiments were performed using the fifth harmonic 212 nm radiation of a separate Nd:YAG laser to photodissociate $CBr_4$ molecules. The use of this wavelength was particularly important for the H-atom product yield experiments due to large interferences from H-atoms produced by $CH_3CHO$ photolysis at 266 nm which overwhelmed the H-atom reactive signal. As the $CH_3CHO$ absorption cross-section is more than 60 times smaller at 212 nm ($5 \times 10^{-22}$ $cm^2$ compared with the 266 nm value of $3 \times 10^{-20}$ $cm^2$) [39] it was much easier to discriminate between the H-atom sources at 212 nm. A detailed description of the method used to disentangle the photolytic and reactive H-atom signals is provided in the supporting information file (SI). The 212 nm beam was generated by first separating the third harmonic 355 nm output of the laser from the residual 532 nm and 1064 nm radiation by two dichroic mirrors with peak reflectivities at 355 nm. The polarisation of the 355 nm beam was then rotated using a half wave plate before mixing it with the residual 532 nm radiation in a type I beta barium borate (BBO) crystal. The 212 nm beam with pulse energies around 7 mJ was separated from the residual 355 nm and 532 nm beams by two dichroic mirrors with peak reflectivities at 212 nm. This beam was steered into the reactor in a similar manner to the 266 nm laser (although the afocal telescope was not employed here). $CBr_4$ vapour was introduced into the reservoir by flowing a small fraction (40-70 sccm) of the main carrier gas flow over solid $CBr_4$ held in a separate flask at room temperature. The gas flow and gas pressure within the flask were carefully controlled to maintain a fixed $CBr_4$ concentration for any single series of experiments. The $CBr_4$ concentration used was estimated to be less than $2.6 \times 10^{13}$ $cm^{-3}$ based on its saturated vapour pressure. At these levels, the C-atom concentration generated was assumed to be identical at all axial positions along the supersonic flow due to the weak $CBr_4$ absorption cross-section at 266 nm. In addition to ground state atomic carbon, previous studies[7] have shown that carbon atoms in the first excited singlet state C($^1$D) are also formed from CBr$_4$ photolysis at this wavelength at relatively low levels. A discussion of the possible effects of the presence of C($^1$D) atoms on the kinetic measurements can be found in an earlier study,[13] while their effects on the present H-atom yield measurements are discussed in the SI file.

In order to study the kinetics of the C + CH$_3$CHO reaction, as the minor reactant, C-atoms were probed by pulsed VUV LIF by exciting the 2s$^2$2p$^2$ $^3$P$_2$ → 2s$^2$ 2p5d $^3$D$_3$° transition at 115.803 nm and by following the resonant fluorescence emission. Radiation at this wavelength was generated in a two-step procedure. First, a Nd:YAG pump laser (532 nm) was used to excite a dye laser operating on LDS 698 dye dissolved in ethanol to produce a tunable beam around 694.8 nm. This beam was directed by a quartz right angled prism into a beta barium borate doubling crystal to produce radiation around 347.4 nm. The beam exiting the crystal containing both the fundamental and doubled laser light was reflected by two dichroic mirrors optimized for reflection around 355 nm to remove the residual fundamental light. Some of the discarded fundamental radiation was collected by a convex lens and focused into a fibre optic cable coupled to a wavemeter, allowing us to continuously monitor the probe laser wavelength and ensure it was correctly tuned to the C-atom transition. Second, the monochromatic 347.4 nm beam was steered towards the reactor by right angled prisms before being focused into a cell attached at the level of the observation axis by a plano-convex lens. This cell, containing 50 Torr of xenon for frequency tripling purposes with 160 Torr of argon added to optimize the phase matching, allowed us to generate tunable VUV radiation around 115.8 nm. During the experiments examining H-atom product formation, tunable radiation around 121.567 nm was generated in a similar manner, with the dye laser operating at 729 nm (dye LDS 722 in ethanol) and a mixture of 210 Torr of krypton and 540 Torr of argon used in the tripling cell. The VUV beam was divergent following third harmonic generation, so a plano-convex MgF$_2$ lens was used as the exit window of the cell to recollimate it. This process had the secondary advantage of leaving the residual UV beam divergent, due to a large difference in the refractive index of MgF$_2$ between VUV and UV wavelengths, thereby reducing the flux of UV radiation that entered the reactor. The residual UV flux reaching the reactor was reduced further by placing a long (75 cm) sidearm between the cell and the reactor while the use of circular diaphragms along the length of the sidearm prevented a large fraction of the scattered UV light from propagating into the reactor. As the sidearm was open to the reactor itself, residual gases (CH$_3$CHO and CBr$_4$ in particular which absorb strongly in the VUV wavelength range) could freely diffuse into it, potentially attenuating the VUV beam. In order to address this issue, this region was continuously flushed by a small flow of argon or nitrogen carrier gas. The VUV beam interacted with the supersonic flow at right angles, with the detector positioned at right angles to the plane containing the supersonic flow and the VUV laser. A solar blind photomultiplier tube (PMT) was employed to detect the VUV LIF emission from C-atoms and H-atoms present in the supersonic flow. In order to protect the PMT from potentially corrosive gases, a lithium fluoride (LiF) window was placed between the reactor and the PMT. The zone between the PMT and the LiF window was continuously evacuated by a dry scroll pump to avoid atmospheric absorption of the VUV radiation. A LiF lens was positioned within this evacuated region to focus the emitted light onto the PMT photocathode. The signal output line of the PMT was connected to a boxcar integrator interfaced to a computer for gating and signal acquisition purposes, while a 500 MHz oscilloscope was used for monitoring, signal optimization and gate positioning. The boxcar system, oscilloscope and both lasers were synchronized by a digital pulse generator operating at a 10 Hz repetition rate. The atomic fluorescence signals were recorded as a function of time between the photolysis and probe lasers. 30 laser shots were recorded at each time delay with more than 100 values of the time delay used to derive the temporal profiles. The signal baseline level (corresponding to an absence of C-atoms, but including scattered light signals and dark noise) was established by firing the probe laser at several negative delay times with respect to the photolysis laser.

Carrier gases Ar (Messer 99.999 %)and $N_2$ (Messer 99.999 %) were regulated by calibrated mass-flow controllers. Other rare gases Xe (Linde 99.999 %) and Kr (Linde 99.999 %) used for frequency tripling, were taken directly from cylinders without further purification. $CH_3CHO$ (> 99 %) was stored in a refrigerator to prevent degradation between experiments.

**3 Theoretical methods**

Electronic structure calculations based on the ground triplet state potential energy surface (PES) of $C_3H_4O$ were performed using ORCA.[40,41] Geometries and energies of the reactants, products, intermediates and transition states (TS) were optimized by density functional theory (DFT) at the M06-2X/aug-cc-pVTZ level of theory.[42] Rotational constants were obtained for all structures, while harmonic vibrational frequency calculations allowed us to derive their zero-point energies (ZPE). Structure visualisation and a harmonic frequency analysis were performed using Avogadro. [43] This also allowed us to verify the nature of the stationary point; a single imaginary frequency indicated the presence of a TS structure while no imaginary frequencies indicated a stable (reactant, product or intermediate) species. TSs were systematically verified to connect to specified minima through an intrinsic reaction coordinate analysis based on the method developed by Morokuma and coworkers. [44] More accurate single point energy calculations were performed for certain critical structures which were found to lie close in energy to the reactant energy level. These calculations, based on the geometries obtained at the M06-2X/aug-cc-pVTZ level were performed using the domain based local pair-natural orbital approximation of the CCSD(T) coupled cluster theory method (DLPNO-CCSD(T)) [45] employing the aug-cc-pVTZ basis set.

## 4 Results

### 4.1 Potential energy surface

Figure 1 shows a schematic representation of the triplet PES for the C + $CH_3CHO$ reaction.

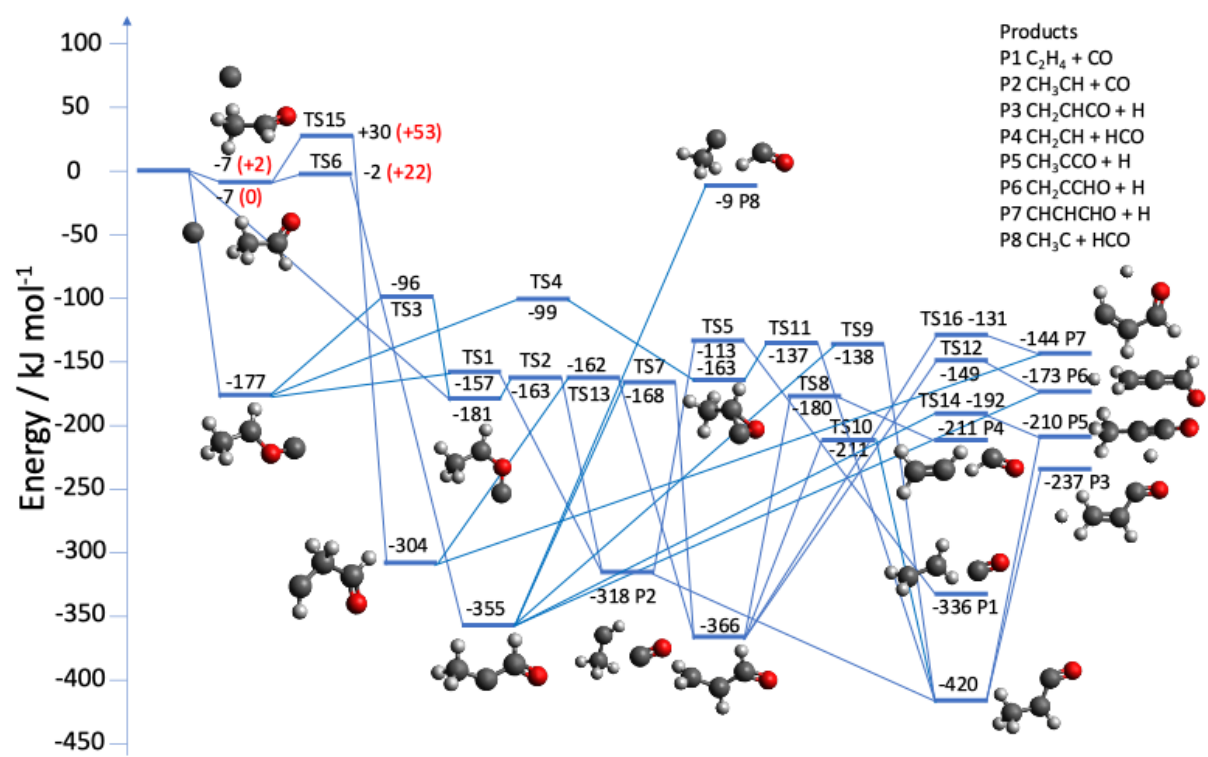

**Figure 1** Triplet potential energy surface for the reaction of C with $CH_3CHO$ with geometries and energies calculated at the M06-2X/aug-cc-pVTZ level (energy values in black). Certain structures close in energy to the reactant asymptote at the M06-2X/aug-cc-pVTZ level were calculated with the more accurate DLPNO-CCSD(T)/ aug-cc-pVTZ method (energy values in red). All energies were corrected for zero-point energy differences.

Nine intermediate species and eight product channels have been identified during the present work, with these species connected by 16 TSs. The calculated harmonic frequencies for these species in addition to their optimized Cartesian coordinates can be found in the SI file. Although there are likely to be other pathways at higher energies, these should not play an important role in the C + $CH_3CHO$ reaction at low temperature. Energies in black have been calculated at the M06-2X/aug-cc-pVTZ level, while energies in red are those calculated by the DLPNO-CCSD(T)/aug-cc-pVTZ method based on the M06-2X/aug-cc-pVTZ level geometries. All values are quoted relative to the separated reactants calculated at the same level and have been corrected for ZPE differences.

Four different entrance channels were identified for this process. The first two channels are characterized by the approach of the carbon atom towards the methyl group of $CH_3CHO$ in two different orientations. The first, via a near colinear approach of the C-atom towards $CH_3CHO$ leads to the formation of a weakly bound van de Waals (vdW) complex calculated to be more stable than the separated reactants by 7 kJ/mol at the DFT level. The second perpendicular approach also results in vdW complex formation, stabilised by 7 kJ/mol at the DFT level. Given their proximity to the reactant energy level, more accurate single point DLPNO-CCSD(T) calculations of these structures were performed, yielding energies of 0 and +2 kJ/mol respectively. The colinear vdW complex could evolve to the $HCCH_2CHO$ intermediate species (-304 kJ/mol) through insertion into a C-H bond of the $CH_3CHO$ methyl group, while the perpendicular vdW complex could evolve to the $CH_3CCHO$ intermediate species (-355 kJ/mol) by insertion into the C-C bond. These pathways are both characterized by TSs, with $HCCH_2CHO$ formation over TS15 (+30 kJ/mol) and $CH_3CCHO$ formation over TS6 (-2 kJ/mol). Given the close proximity of these TSs to the reactant energy level (TS6 in particular), more precise energies of +53 kJ/mol and +22 kJ/mol were derived at the DLPNO-CCSD(T)/aug-cc-pVTZ level of theory for TS15 and TS6 respectively. Consequently, given the low stability of these complexes coupled with the high barriers to isomerization, it seems unlikely that these pathways play any significant role in the reaction of C with $CH_3CHO$ at room temperature and below. Although tunnelling could conceivably play a role for this system, previous studies of related reactions[11] have shown the requirement for a relatively stable pre-reactive complex and a low energy barrier for tunnelling to be efficient. In the present case, although the isomerization barriers are relatively low, the absence of a stable pre-reactive complex almost certainly precludes the possibility for reaction to occur through tunnelling.

The third and fourth reaction pathways are characterized by approach of the carbon atom towards the oxygen atom of the aldehyde functional group, forming two different rotational isomers with the formula $CH_3CHOC$ with energies of -177 kJ/mol ($CH_3CHOC$-1) and -181 kJ/mol ($CH_3CHOC$-2) clearly corresponding to the direct formation of a chemical bond. These two species can interconvert over TS3, 96 kJ/mol below the reactants while $CH_3CHOC$-1 can also evolve to a cyclic species (c-$CH_3CHOC$, -163 kJ/mol) over TS4, 99 kJ/mol below the reactants. By dissociation of the original C-O bond of the $CH_3CHO$ molecule, both $CH_3CHOC$ intermediates can form the product species $CH_3CH$ + CO (P2) via TS1 ($CH_3CHOC$-1 – P2) and TS2 ($CH_3CHOC$-2 – P2) only 20 kJ/mol and 18 kJ/mol above the respective $CH_3CHOC$ intermediate. These products can evolve further, notably through the isomerization of $CH_3CH$ to $^3C_2H_4$ over TS5 (-113 kJ/mol) or by reaction between the products $CH_3CH$ and CO themselves through C-C bond formation, leading to the barrierless formation of a highly stable intermediate species $CH_3CHCO$ (-420 kJ/mol). The two C-H bond fission steps of $CH_3CHCO$ dissociation are also barrierless, yielding bimolecular products P3, $CH_2CHCO$ +H (-237 kJ/mol) and P5, $CH_3CCO$ + H (-210 kJ/mol). The present calculations show that once $CH_3CHCO$ has been formed, other intermediate species such as $CH_2CHCHO$, $CH_3CCHO$ and $HCCH_2CHO$ are also potentially accessible through the various isomerizations brought about by H-atom transfer. As these processes involve a range of TSs with energies between -211 and -138 kJ/mol, they are all potentially relevant at low temperature and open up the possibility of alternative bimolecular products such as $CH_2CH$ + HCO (P4, -211 kJ/mol) from C-C bond fission in $CH_2CHCHO$, while $CH_2CCHO$ + H (P6, -173 kJ/mol) and $CHCHCHO$ + H (P7, -144 kJ/mol) can be accessed by the various C-H bond dissociations of the $CH_2CHCHO$, $CH_3CCHO$ and $HCCH_2CHO$ intermediates. Nevertheless, considering the low energy of the pathways from C+$CH_3CHO$ → $CH_3CHOC$ → P2 with only low barriers for the formation of $CH_3C$ + CO, and the further evolution of P2 → $CH_3CHCO$ → P3 & P5 it seems likely that $CH_3C$ + CO represent the major products of the C + $CH_3CHO$ reaction, with possible minor contributions from the H + $CH_2CHCO$ and H + $CH_3CCO$ products following $CH_3CHCO$ dissociation. As a general comment on the nature of the reaction based on the calculated PES, given the barrierless nature of the entrance channel pathways leading to the two $CH_3CHOC$ intermediates coupled with the absence of barriers close to the reactant level, we would expect the reaction to be rapid at low temperature with little or no temperature dependence. Furthermore, we would not expect the reaction to lead to large H-atom yields, as formation of the products $CH_3CH/C_2H_4$ + CO is much more thermodynamically favourable. In our earlier study of the C + acetone ($CH_3COCH_3$) reaction[13] we concluded that intersystem crossing might play a role in the transformation of the $^3CH_3CCH_3$ products to the singlet state followed by isomerisation to propene, $CH_2CHCH_3$. In a similar manner, the $^3CH_3CH$ or $^3C_2H_4$ products formed by the C + $CH_3CHO$ reaction could eventually lead to the formation of ethene in the ground singlet state, $^1C_2H_4$. Indeed, a minimum energy crossing point calculation performed at the M06-2X/aug-cc-pVTZ level of theory leads to an energy difference value close to zero between the triplet and singlet states, so intersystem crossing could be quite favourable in this instance.

**4.2 Rate constants**

In order to simplify the rate constant analysis, the isolation method was applied for all the measurements described here. In this respect, $CH_3CHO$ was chosen as the excess reactant so that its concentration did not change during the course of the experiments. As the minor reactant species, the fluorescence emission from unreacted C-atoms within the supersonic flow varied exponentially as a function of time obeying a pseudo-first-order rate law of the form

$$I_C(t) = I_{C_0} \exp(-k_{1st}t) \qquad (E1)$$

to yield C-atom temporal profiles similar to those shown in Figure 2.

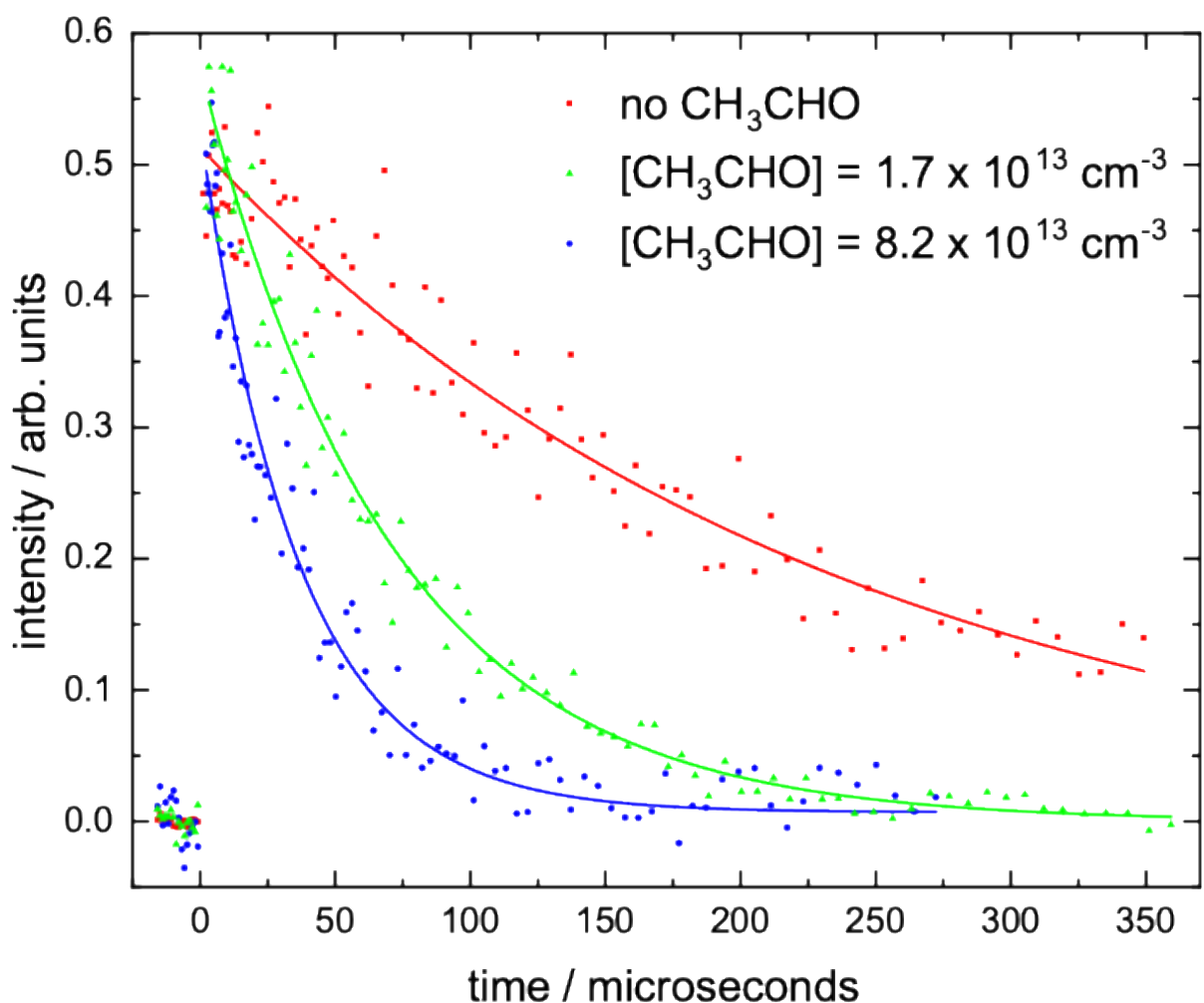

**Figure 2** Atomic carbon VUV LIF intensity profiles recorded at 50 K as a function of the time between photolysis laser and probe laser pulses. (Red squares) without CH$_3$CHO; (green triangles) [CH$_3$CHO] = 1.7 × 10$^{13}$ cm$^{-3}$; (blue circles) [CH$_3$CHO] = 8.2 × 10$^{13}$ cm$^{-3}$. Solid lines represent non-linear least squares exponential fits to the individual datasets based on expression (E1).

Here, $I_{C_0}$ and $I_C(t)$ are the initial and time dependent C-atom VUV LIF signals respectively and $k_{1st}$ is the pseudo-first-order rate constant, equal to $k_{C+CH_3CHO}[CH_3CHO]$ assuming no secondary losses of atomic carbon. In reality, carbon atoms are also lost through secondary reactions such as those with the C-atom precursor molecule CBr$_4$ which is also present in excess and by diffusion out of the detection region as shown by the red C-atom temporal profile in Figure 2 where no acetaldehyde molecules were present. In this respect, for these experiments $k_{1st}$ is given by the expression

$$k_{1st} = k_{C+CH_3CHO}[CH_3CHO] + k_{C+CBr_4}[CBr_4] + k_{diff} \qquad (E2)$$

As the CBr$_4$ concentration is not varied for any single series of measurements, the contributions of the second and third terms of this expression to the overall pseudo-first-order rate constant are constant so that the observed variation in $k_{1st}$ between the green and blue curves shown in Figure 2 is entirely due to the changing value of the first term of expression (E2). At least 24 decay profiles were recorded at each temperature, for at least five different values of [CH$_3$CHO]. The $k_{1st}$ values derived from exponential fits to these decay profiles were plotted as a function of [CH$_3$CHO], allowing us to determine the second-order rate constant at a given temperature from the gradients of weighted linear least-squares fits to the data sets. Each datapoint was weighted according to the formula $1/\sigma^2$ where $\sigma$ is the standard deviation attributed to the decay constant derived during the exponential fitting procedure. This ensured that the data points with the largest uncertainties were also those with the smallest weight. Several of these second-order plots are shown in Figure 3. Unless otherwise mentioned, the C-atoms used in these experiments were produced by CBr$_4$ photolysis at 266 nm.

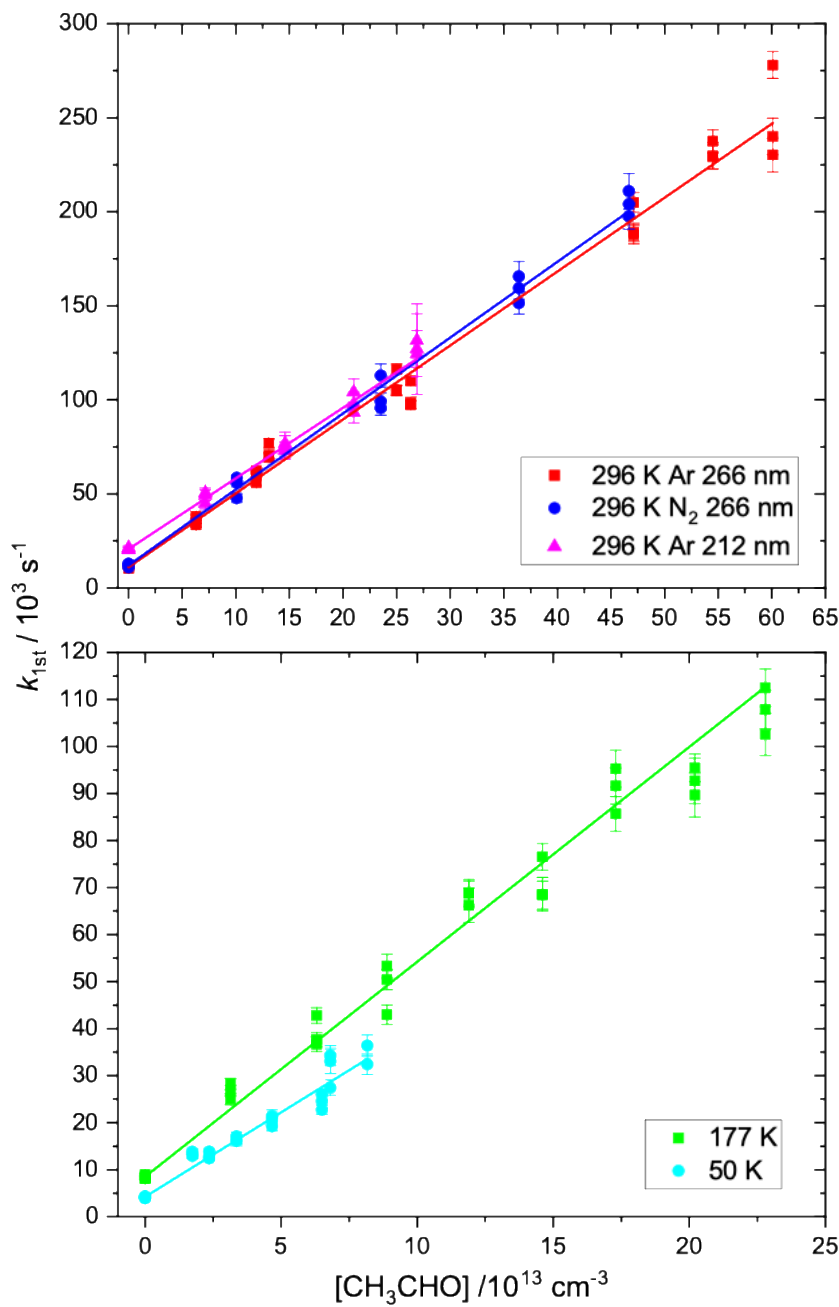

**Figure 3** Plots of the pseudo-first-order rate constant as a function of the CH₃CHO concentration at selected temperatures 296 K, 177 K and 50 K. Top panel: data recorded at 296 K. (red squares) argon carrier gas; (blue circles) nitrogen carrier gas; (purple triangles) argon carrier gas, C-atoms produced by 212 nm photolysis. Bottom panel: data recorded at 177 K and 50 K. (green squares) 177 K data with nitrogen carrier gas; (cyan circles) 50 K data with argon carrier gas. Solid lines represent weighted linear least-squares fits to the data where the gradient yields the second-order rate constant. Error bars represent the uncertainties of individual pseudo-first-order rate constants derived from non-linear fits to the atomic carbon fluorescence profiles such as those shown in Figure 2.

It can be seen from Figure 3 that the range of [CH₃CHO] used at 50 K (and at 75 K) is much smaller than those used at higher temperatures. This restricted concentration range was used due to the likely formation of CH₃CHO clusters at higher concentration levels, leading to pseudo-first-order rate constants lower than expected from second order fits to the data obtained at lower [CH₃CHO]. Consequently, the data recorded at higher [CH₃CHO] were excluded from the fits. The y-axis intercept values of the plots shown in Figure 3 correspond to the sum of the second and third terms of expression (E2). Interestingly, it can be seen that the y-axis intercept value for the room temperature experiments employing CBr₄ photolysis at 212 nm is approximately twice as large as those of the corresponding experiments employing CBr₄ photolysis at 266 nm (average decay constants of approximately 21000 and 11000 s⁻¹ respectively). As the signal levels were considerably lower for the experiments conducted at 212 nm, the CBr₄ concentration used during these experiments was 40 % larger than the one used with 266 nm photolysis. Additionally, there may also have been an increase of the diffusion term contribution with 212 nm photodissociation of CBr₄ due to the higher initial energy shared between the fragments. The measured second-order rate constants are summarized in Table 1 alongside other information such as the number of measurements, the acetaldehyde concentration ranges and the carrier gases used. The rate constants are also plotted as a function of temperature in Figure 4 alongside previous work.

**Table 1** Derived second-order rate constants for the C($^3$P) + CH$_3$CHO reaction

| T / K | $N^b$ | [CH₃CHO] / 10¹³ cm⁻³ | Flow density] / 10¹⁷ cm⁻³ | $k_{C(^3P)+CH_3CHO}$ / 10⁻¹⁰ cm³ s⁻¹ | Carrier gas |
|---|---|---|---|---|---|
| 296 | 27 | 0 - 54.5 | 1.65 | (3.93 ± 0.40)$^c$ | Ar |
| 296 | 15 | 0 - 46.7 | 1.65 | (4.03 ± 0.42)$^c$ | N₂ |
| 296 | 15 | 0 - 26.9 | 1.65 | (3.76 ± 0.38)$^c$ | Ar (212 nm) |
| 177 ± 2$^a$ | 30 | 0 - 22.8 | 0.94 | (4.57 ± 0.47) | N₂ |
| 127 ± 2 | 30 | 0 - 33.4 | 1.26 | (4.03 ± 0.41) | Ar |
| 75 ± 2 | 24 | 0 - 6.1 | 1.47 | (4.29 ± 0.47) | Ar |
| 50 ± 1 | 26 | 0 - 8.2 | 2.59 | (3.59 ± 0.38) | Ar |

[a] Uncertainties on the calculated temperatures represent the statistical (1σ) errors obtained from Pitot tube measurements of the impact pressure. [b] Number of individual measurements. [c] Uncertainties on the measured rate constants represent the combined statistical (1σ) and estimated systematic errors (10%).

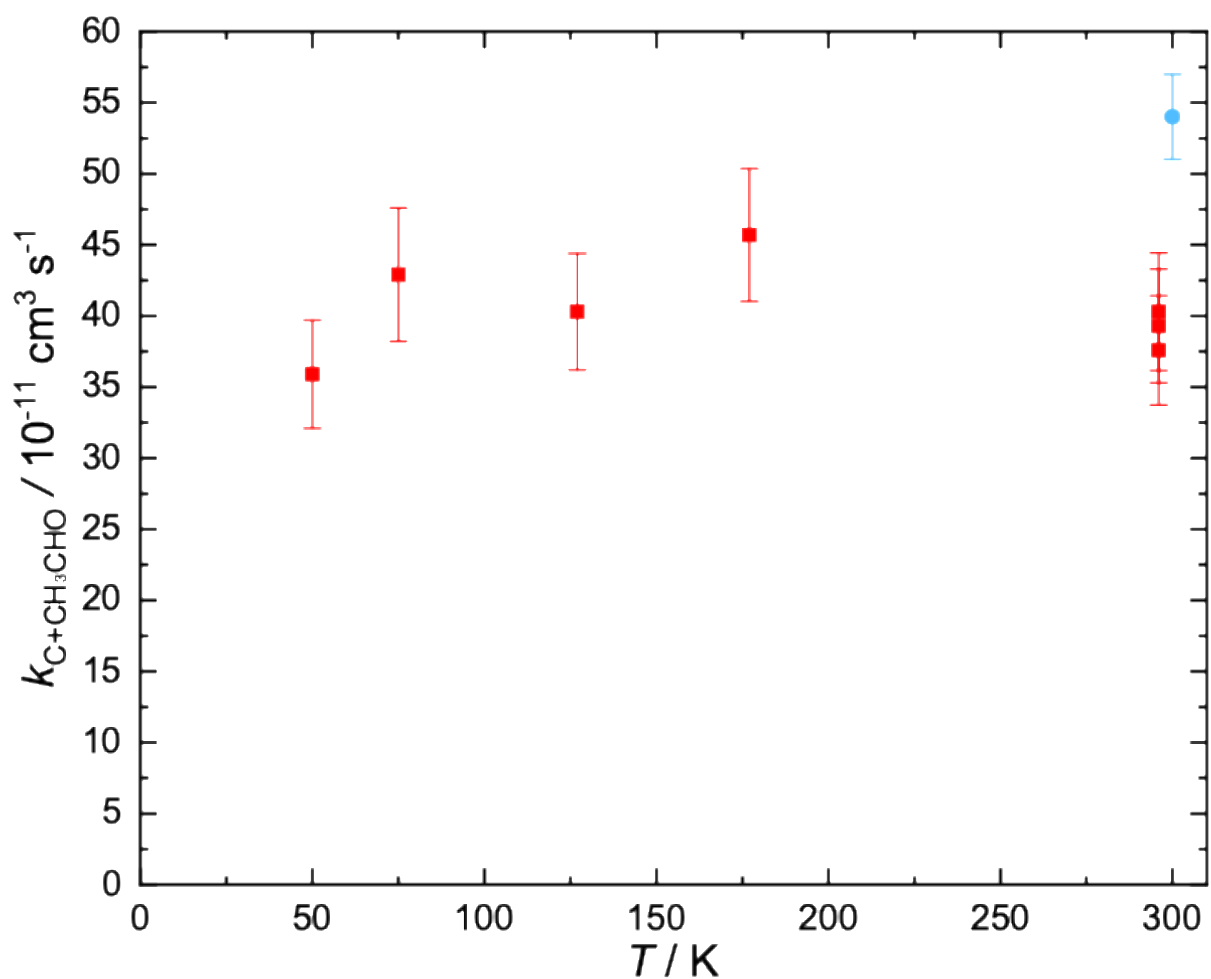

**Figure 4** Measured second-order rate constants for the C + CH$_3$CHO reaction as a function of temperature. (Red squares) this work; (blue circle) Husain and Ioannou.[27] Error bars on the values measured during this work represent the combined statistical and systematic uncertainties. Statistical uncertainties were derived from weighted linear-least squares fits such as those shown in Figure 3. Systematic uncertainties were estimated to be 10 % of the nominal value of the second-order rate constant.

The derived second-order rate constants are all very large, with values ranging from (3.6-4.6) × $10^{-10}$ cm$^3$ s$^{-1}$ over the 50-296 K temperature range. These findings indicate that there is at least one barrierless pathway for reaction to occur, in good agreement with the results of our electronic structure calculations shown in Figure 1. Indeed, the absence of a noticeable temperature dependence for the measured rate constants clearly supports the hypothesis that the reaction involves the formation of a stable CH$_3$CHOC intermediate species in the entrance channel (rather than the formation of the two other weakly bound complexes), followed by direct dissociation or isomerization processes involving only very low barriers with respect to the reactant energy level. As the rate constants are mostly within the combined measurement uncertainties, we recommend the use of a temperature independent value for the rate constant, equal to $k_{\text{C+CH}_3\text{CHO}} = (4.0 \pm 0.8) \times 10^{-10}$ cm$^3$ s$^{-1}$, derived from an average of the individual second-order rate constants determined here. The error bars on this value represent the derived statistical uncertainty with a 10 % systematic uncertainty added.

There is only a single previous study of the kinetics of the C + CH$_3$CHO reaction, by Husain and Ioannou[27] who determined a rate constant of = (5.4 ± 0.3) × $10^{-10}$ cm$^3$ s$^{-1}$. These authors employed a coaxial lamp to produce C-atoms in their work through the VUV photolysis of carbon suboxide, C$_3$O$_2$, at wavelengths greater than 160 nm. Ground state carbon atoms in the presence of a range of aldehydes and ketones were followed by absorption spectroscopy using an atomic resonance lamp coupled with a VUV monochromator to optically isolate the strong atomic carbon transitions around 166 nm. A solar blind PMT was used to detect the transmitted light around this wavelength. The rate constant derived by Husain and Ioannou of (5.4 ± 0.3) × $10^{-10}$ cm$^3$ s$^{-1}$ at room temperature is approximately 25 % larger than the values obtained in the present work and outside of the combined error bars. This discrepancy could originate from several sources including the use of a much less sensitive detection method by Husain and Ioannou which would have required the use of much larger C-atom concentrations during the experiments. Moreover, measurements were performed using reactants (CH$_3$CHO), precursors (C$_3$O$_2$) and buffer gases (He) that were mixed and stored in bulbs prior to the experiments, thereby increasing the possibility of surface reactions and additional impurities. Consequently, there was likely to be an increased potential for interferences from secondary reactions in these earlier experiments.

**4.3 H-atom yields**

In order to provide some information about the possible exit channels of the C + CH$_3$CHO reaction we also studied the formation of atomic hydrogen products by recording the VUV LIF intensities as a function of time following excitation at 121.567 nm. These relative intensities were calibrated by comparison with those recorded for a reference process, namely the C + C$_2$H$_4$ → C$_3$H$_3$ + H reaction. The H-atom yield of the C + C$_2$H$_4$ reaction has already been determined experimentally with a value of 0.92 ± 0.04 at 300 K,[46] allowing the measured relative intensities for the C + CH$_3$CHO reaction to be converted to absolute ones. The H-atom yield of the C + C$_2$H$_4$ reaction was assumed not to vary as a function of temperature as explained in previous work.[13]

Preliminary experiments using the 266 nm photolysis of CBr$_4$ as the source of carbon atoms indicated that CH$_3$CHO was readily photodissociated at this wavelength, producing large quantities of H-atoms in the reactor. The CH$_3$CHO absorption cross-section at 266 nm has been measured with a value of 3.1 × 10$^{-20}$ cm$^2$.[39] One of the experiments we performed at 296 K with the 266 nm photolysis laser is shown in Figure S1 of the SI file, alongside the equivalent experiment performed with the 212 nm photolysis laser where the CH$_3$CHO absorption cross-section = 4.9 × 10$^{-22}$ cm$^2$.[39] The data obtained by photolysis at 212 nm was considered to be more reliable than the data obtained by photolysis at 266 nm, and following the data analysis steps described in the SI, we obtain the H-atom production curves for the target C + CH$_3$CHO and reference C + C$_2$H$_4$ reactions shown in Figure 5, using a CBr$_4$ photolysis wavelength of 212nm.

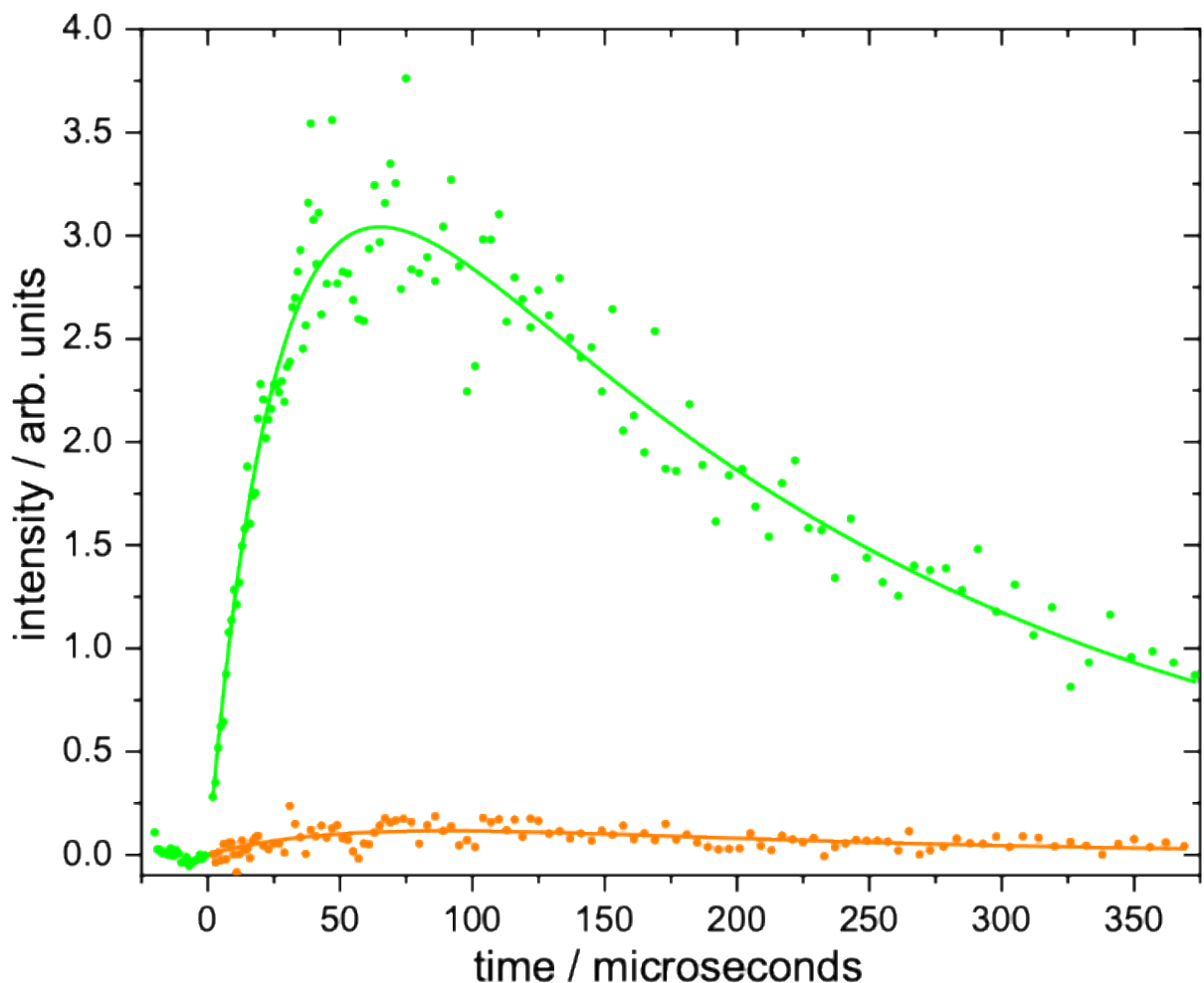

**Figure 5** H-atom fluorescence profiles as a function of reaction time recorded at 296 K for experiments performed with 212 nm photolysis. (Green circles) H-atom signal from the C + $C_2H_4$ reference reaction with $[C_2H_4]$ = 4.1 × $10^{13}$ $cm^{-3}$; (orange circles) corrected H-atom signal from the C + $CH_3CHO$ reaction with $[CH_3CHO]$ = 2.2 × $10^{13}$ $cm^{-3}$ (see SI file for details). The solid lines represent biexponential fits to the corresponding data sets using expression (E3).

Both the green and the orange data are well described by the expected biexponential expression

$$I_H = A\{\exp(-k_{L(H)}t) - \exp(-k_{1st}t)\} \qquad (E3)$$

where $I_H$ is the time dependent H-atom intensity, $k_{L(H)}$ is the rate constant for H-atom loss, $k_{1st}$ is the rate constant for H-atom formation and A is the H-atom signal amplitude.

If we divide the peak amplitude of the fit to the reactive signal by the peak amplitude of the reference signal fit, and correct for the fact that the absolute H-atom yield of the reference reaction is 0.92, [46] we obtain a value for the H-atom yield of the C + $CH_3CHO$ reaction of 3.5 % at 296 K while the average value derived from seven experiments of this type was (3.3 ± 1.1) %. No further corrections, such as those potentially arising from absorption of the 121.567 nm excitation laser or fluorescence by $CH_3CHO$ or $C_2H_4$ were required due to the low coreactant concentration levels used.

When we use the same type of analysis for experiments performed at 177 K, to obtain the equivalent H-atom yields at this temperature, curves of the type shown in Figure 6 are obtained.

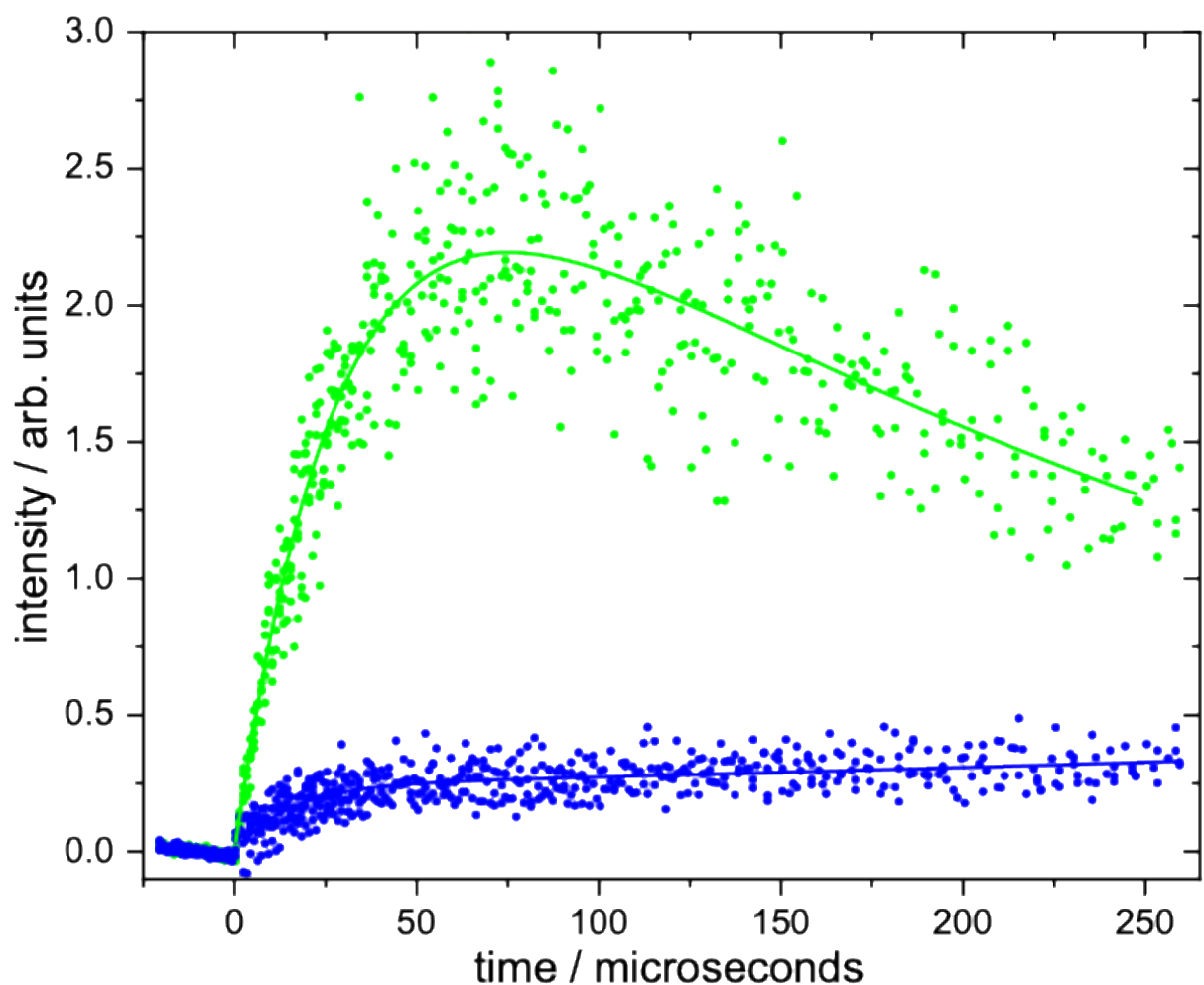

**Figure 6** H-atom fluorescence profiles as a function of reaction time recorded at 177 K for experiments performed with 212 nm photolysis. (Blue circles) corrected H-atom signal from the C + $CH_3CHO$ reaction with [$CH_3CHO$] = 2.9 × $10^{13}$ $cm^{-3}$ (see SI file for details); (green circles) H-atom signal from the C + $C_2H_4$ reference reaction with [$C_2H_4$] = 2.7 × $10^{13}$ $cm^{-3}$.

Surprisingly, the H-atom formation curve from the C + CH$_3$CHO reaction represented by the blue data points in Figure 6 does not display the expected behaviour as the H-atom VUV LIF signal continues to increase with time, corresponding to an unrealistic negative value of the time constant $k_{L(H)}$ for the first term of expression (E3). This behaviour was never observed during any of the experiments conducted at room temperature. As we do not have a good handle on the magnitude of the secondary H-atom signal at 177 K, it is difficult to provide a precise value for the H-atom yield at this temperature. Nevertheless, if we take the values of the fits to the reactive signal curves at times corresponding to the peak of the reference signal curve (around 75 μs for the example given in Figure 7) we obtain H-atom yields in the range 7-15 % for all the experiments we performed with an average value of around 11 % which will clearly overestimate the real value due to the secondary H-atom contribution. There could be several possible sources for the additional H-atom products observed in these experiments. Secondary reactions between one of the primary products of the C + CH$_3$CHO reaction with CH$_3$CHO itself could lead to the formation of additional H-atoms. In this respect, if we examine the reactive triplet PES in Figure 1, CH$_3$CH is likely to be a major product which could go on to react with CH$_3$CHO. Despite this, the related $^3$CH$_2$ species has been shown to react slowly with CH$_3$CHO at room temperature and above.[47] Another possibility to account for the additional H-atom formation could be the decomposition of one or more of the primary reaction products such as HCO or CH$_3$CH. Product HCO dissociation was hypothesized to be a source of atomic hydrogen in our earlier work on the C + CH$_3$OH reaction,[7] while electronic structure calculations performed at the M06-2X/aug-cc-pVTZ level in the present work indicate that the dissociation of CH$_3$CH to CH$_2$CH + H is a barrierless process, with the CH$_2$CH + H products 211 kJ/mol below the C + CH$_3$CHO level according to Figure 1. Although both of these possibilities could lead to H-atom production at 177 K, it is unclear why we do not observe a similar phenomenon at 296 K. Considering the observed discrepancies in the reactive H-atom signal profiles obtained at different temperatures, we are unable to provide a quantitative value for the H-atom yields for the C + CH$_3$CHO reaction at the present time. Despite this, on a qualitative level, the channels leading to H-atom formation are clearly very minor with values almost certainly considerably lower than 10 %. Considering the theoretical analysis presented in section 4.1, the major reaction products are likely to be either CH$_3$CH + CO or C$_2$H$_4$ + CO.

**5 Astrochemical model**

As already indicated in section 1, $CH_3CHO$ can be produced by various processes, the $O + C_2H_5$ reaction[48-51] being currently the most important gas-phase source of $CH_3CHO$ according to astrochemical models based on the kida.uva.2024 network.[52] Indeed, the ionic pathway initiated by the radiative association between $H_3O^+$ and $C_2H_2$ proposed by Herbst[53] as a source of $CH_3CHO$ subsequently proved less effective than expected[54-56] and currently plays a secondary role. It should be noted that the production of the $C_2H_5$ radical is not obvious in current models because the ionic route, the dissociative recombination (DR) of $C_2H_6^+$ and $C_2H_7^+$, is ineffective because the hydrogenation of $C_2H_x^+$ stops at $C_2H_4^+$. We have therefore re-examined the formation of $C_2H_5$ and $CH_3CHO$ in the gas phase and tested the effects induced by including the $C + CH_3CHO$ reaction in a standard astrochemical model. The Nautilus code[57] was used to perform the present simulations. Nautilus is a 3-phase (gas, dust grain ice surface and dust grain ice mantle) time dependent chemical model employing kida.uva.2014[52] as the basic reaction network. This network, with 800 individual species involved in approximately 9000 separate reactions, was recently updated for a better description of COM chemistry on interstellar dust grains and in the gas-phase.[58, 59] Elements are either initially in their atomic or ionic forms in this model except for $H_2$ (elements with an ionization potential < 13.6 eV are considered to be fully ionized) and the C/O elemental ratio is set to 0.71 in this work. The initial simulation parameters are identical to the ones used in Hickson, Loison and Wakelam[14] (see Table 4 of that paper). The grain surface and the mantle are both chemically active for these simulations, while accretion and desorption are only allowed between the surface and the gas-phase. The dust-to-gas ratio (in terms of mass) is 0.01. A sticking probability of 1 is assumed for all neutral species while desorption occurs by both thermal and non-thermal processes (cosmic rays, chemical desorption) including sputtering of ices by cosmic-ray collisions.[60]

Figure 7 shows the reactions involved in $CH_3CHO$ production (top panel) and destruction (bottom panel) with the individual reaction fluxes (flux = rate constant × species 1 abundance × species 2 abundance) plotted as a function of cloud age.

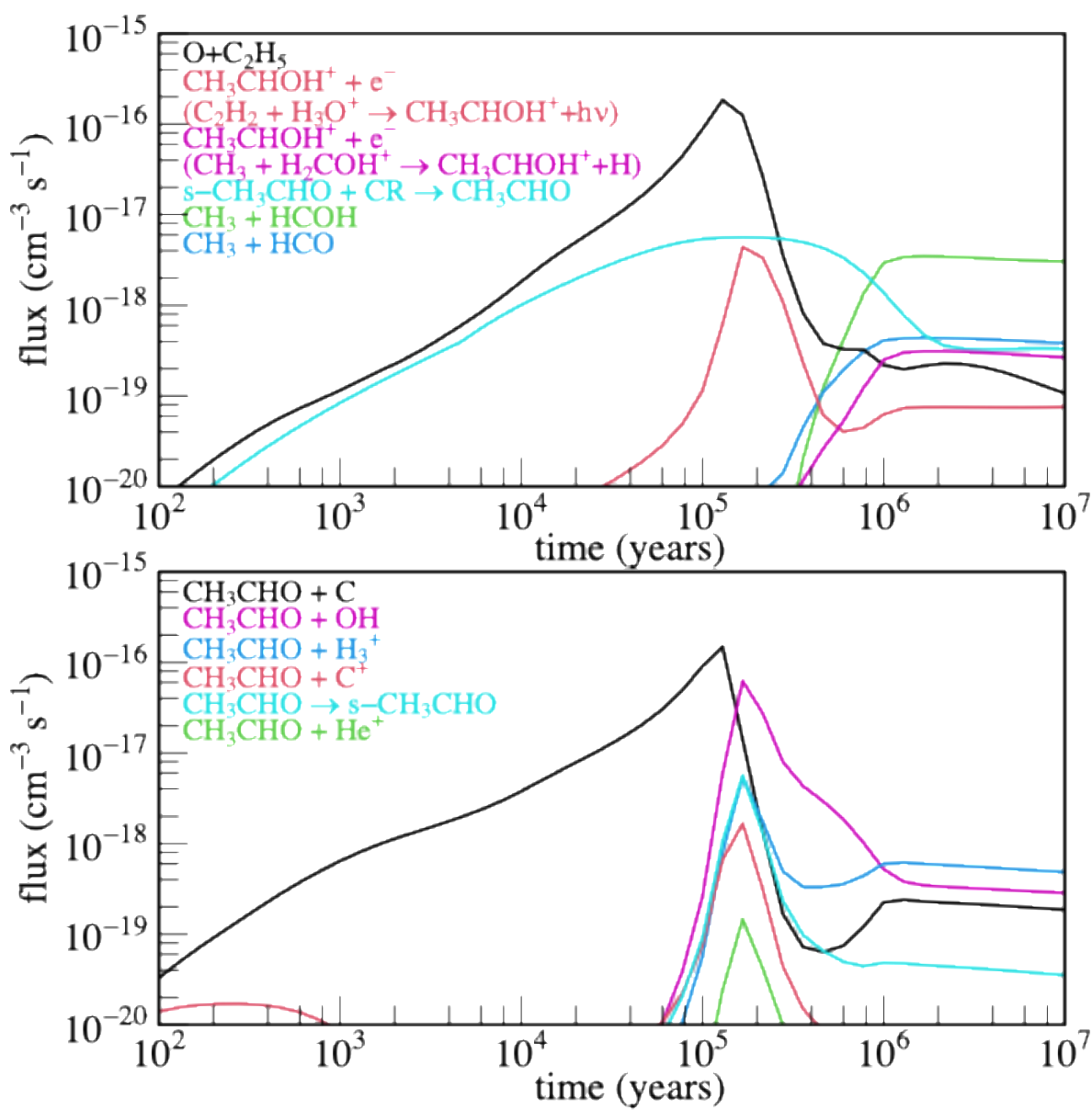

**Figure 7** Simulated reaction fluxes obtained by the astrochemical model Nautilus. (Top panel) Major CH₃CHO formation pathways as a function of cloud age. (Bottom panel) Major CH₃CHO destruction pathways as a function of cloud age.

In our updated network various reactions lead to CH₃CHO formation. The most efficient one is the O + C₂H₅ reaction,[48-51] followed by the radiative association between H₃O⁺ and C₂H₂ to give CH₃CHOH⁺ which undergoes DR to form CH₃CHO. The third most efficient gas-phase CH₃CHO production mechanism is desorption of s-CH₃CHO (s- for grain surface) by cosmic-ray collisions.[60] This process is non negligible because s-CH₃CHO is efficiently produced through the s-C + s-CO → s-C₂O reaction, (a barrierless reaction in the gas-phase[61] that is also likely to be the case on grain surfaces), followed by hydrogenation of s-C₂O. In our updated network

S24

we have introduced a second product channel for the gas-phase $CH_3$ + HCO reaction in addition to the $CH_4$ + CO one, namely, radiative association to give $CH_3CHO$. This addition channel is well known experimentally and has been measured to be equal to 30-50% of the total rate at high pressure.[62, 63] Another new $CH_3CHO$ production pathway is the $CH_3$ + HCOH reaction. HCOH is an isomer of $H_2CO$, produced efficiently through the $CH_3$ + OH reaction studied theoretically[64, 65] and which should be studied in more detail in the future. We have also made a preliminary theoretical study of the reaction $CH_3 + H_2COH^+ \rightarrow CH_3CHOH^+ + H$ which is exothermic by 52 kJ/mol at the M06-2X/AVTZ level and therefore a potential source of $CH_3CHO$ through the DR of $CH_3CHOH^+$.[66] Even if they play a significant role, the three new reactions introduced into our model only play a role at times beyond the chemical age determined by the observational-modelling agreement,[67] around 1-2 × $10^5$ years with the current model. It should be noted that with our chemical network, $C_2H_5$ (the main precursor of $CH_3CHO$) is mainly produced by the reaction between O and $C_3H_6$. This reaction has been studied experimentally at low temperature (23-298 K), where it remains efficient through tunneling,[68] and forms $C_2H_5$ + HCO among other products.[69] So even if the formation of $C_3H_6$ is ambiguous in dense molecular clouds,[70] its high observed abundance[71] makes the production pathway for $C_2H_5$, and therefore $CH_3CHO$, reasonably well described.

The main losses of gas-phase $CH_3CHO$ are through its reactions with atomic C (this work), with the OH radical (measured down to 11.7 K[72]), with ions ($H_3^+$, $C^+$, $He^+$ and $HCO^+$ (not shown in Figure 7)) and by depletion onto grains. Note that the OH + $CH_3CHO$ reaction involves a relatively large flux and is therefore a non-negligible source of $CH_3CO$ radicals in dense molecular clouds.[73]

The dashed lines in Figure 8 show the simulated abundances of several species including $CH_3CHO$ generated by the model in the presence (solid lines) and absence (dashed lines) of the C + $CH_3CHO$ reaction.

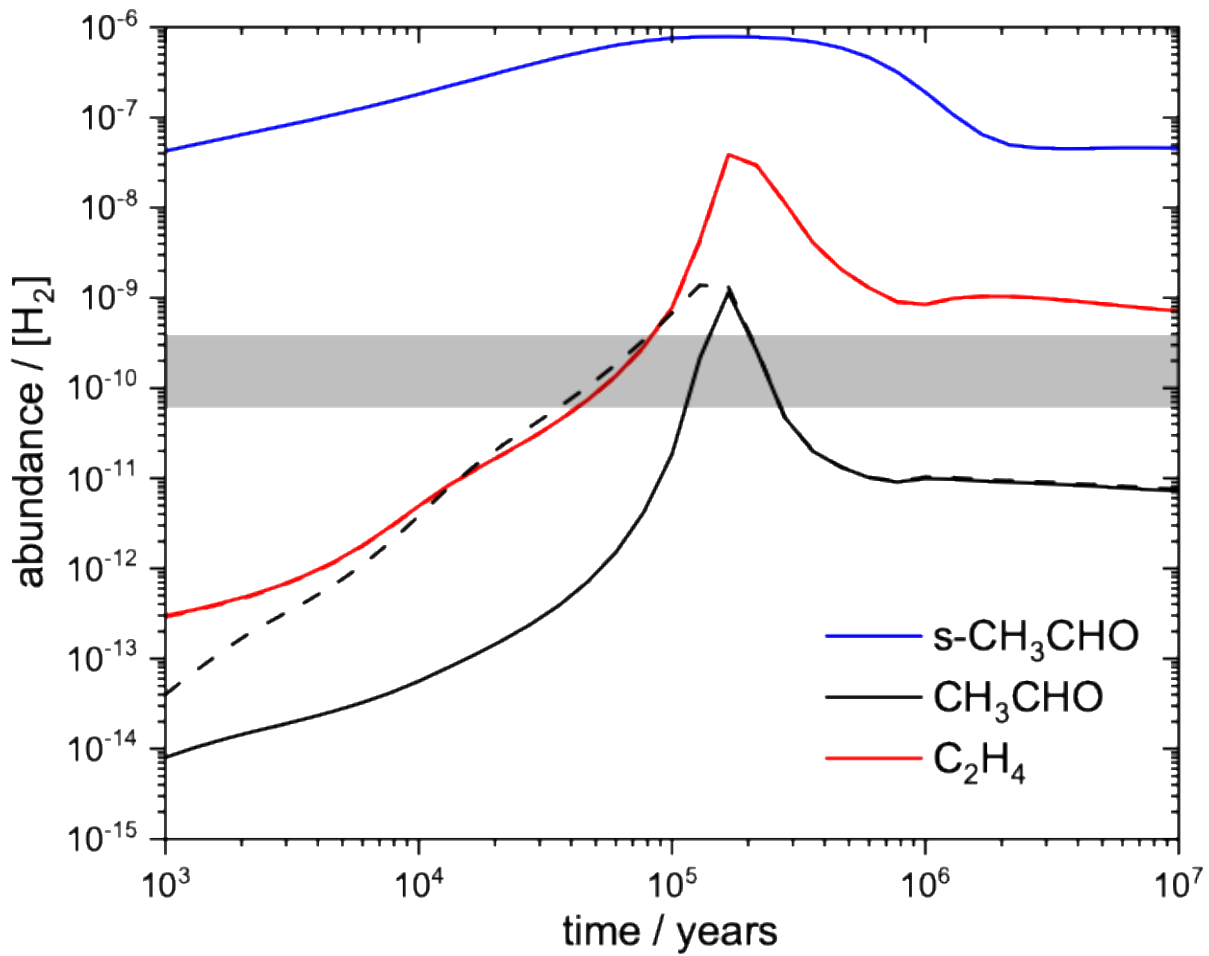

**Figure 8** Gas-grain astrochemical model results for the formation of gas-phase acetaldehyde, CH$_3$CHO (black lines), acetaldehyde on interstellar ice grains, s-CH$_3$CHO (blue lines) and ethene, C$_2$H$_4$ (red lines). (Dashed lines) standard network results. (Solid lines) standard network with the C + CH$_3$CHO reaction added. Dashed and solid lines overlap for s-CH$_3$CHO and C$_2$H$_4$. The horizontal grey rectangle represents the range of observed CH$_3$CHO abundances in Taurus Molecular Cloud. [21]

The inclusion of the C + CH$_3$CHO reaction in the network with an estimated rate constant, $k_{\text{C+CH}_3\text{CHO}}(10\text{K}) = 4.0 \times 10^{-10}$ cm$^3$ s$^{-1}$ leads to a maximum decrease of the CH$_3$CHO abundance of more than an order of magnitude between 10$^3$ and 10$^5$ years. For astrochemical modeling purposes, C$_2$H$_4$ + CO were considered to be the sole products of the C + CH$_3$CHO reaction for these simulations, based on the experimental and theoretical results presented above. Moreover, the isomer CH$_3$CH was not considered here as this species is not currently included in the reaction network. At ages considered to be characteristic of typical dense clouds such as TMC-1 (a few 10$^5$ years), atomic carbon has been mostly removed from the gas-phase through reactions forming CO and by accretion onto grains, thereby limiting the effect of the C + $CH_3CHO$ reaction at longer times. Despite the large effect of the C + $CH_3CHO$ reaction on $CH_3CHO$ abundances at early times, these simulations indicate that the C + $CH_3CHO$ reaction induces only small changes in the gas-phase $CH_3CHO$ abundance at typical dense interstellar cloud ages, with a calculated $CH_3CHO$ abundance that reproduces reasonably well the observed one for TMC-1 at these times. As a major product of the C + $CH_3CHO$ reaction it is also interesting to examine the effect of including this reaction in the network on $C_2H_4$ abundances. It can be seen from Figure 8 that the $C_2H_4$ abundance is mostly unaffected by the C + $CH_3CHO$ reaction, where the simulations with and without this reaction (solid and dashed red lines) are almost identical.

## 6 Conclusions

This paper describes the results of an experimental and theoretical investigation of the C + $CH_3CHO$ reaction in the gas-phase, coupled with an astrochemical modeling study of the influence of this process in dense interstellar clouds. On the experimental side, a supersonic flow reactor was used to determine the rate constants and qualitative H-atom yields of this reaction over the 50-296 K temperature range using a pulsed laser photolysis-laser induced fluorescence method to generate and detect atomic carbon (atomic hydrogen products were also detected by laser induced fluorescence). In order to help with our interpretation of the experimental results, calculations of the reactive triplet potential energy surface were also performed. The electronic structure calculations predict no or low energy barriers, in good agreement with the large measured rate constants and weak temperature dependence of the reaction rate. In addition, the theoretical results also support the observation of very low H-atom yields by indicating the barrierless nature of the pathways leading to $^3CH_3CH/^3C_2H_4$ + CO as the major products at all temperatures, with possible further evolution of $^3C_2H_4$ to $^1C_2H_4$ through intersystem crossing. Although the astrochemical simulations clearly show the important role of the C + $CH_3CHO$ reaction in regulating the $CH_3CHO$ abundance in less evolved molecular clouds, the reaction has little influence on $CH_3CHO$ abundance levels in more evolved clouds and is only a minor source of $C_2H_4$ production throughout the entire cloud lifetime.

**Author Information**

**Corresponding Author**

*Email: kevin.hickson@u-bordeaux.fr.

**Supporting Information**

Supporting figures (Figures S1-S3) and methodology for the H-atom yield analysis, frequencies (Tables S1-S3) and geometries (Tables S4-S6) of the stationary points involved in the C($^3$P) + CH$_3$CHO reaction obtained at the M06-2X/aug-cc-pVTZ level of theory.

**Acknowledgements**

K. M. H. and V. W. acknowledge support by the thematic action "Physique et Chimie du Milieu Interstellaire" (PCMI) of the INSU National Programme "Astro", with contributions from CNRS Physique & CNRS Chimie, CEA, and CNES. K. M. H. also acknowledges support from the ''Programme National de Planétologie'' (PNP) of the CNRS/INSU.

Supporting information file for

**Rate constants and product yields for the C + CH$_3$CHO reaction at low temperatures**

Kevin M. Hickson,[1,*] Jean-Christophe Loison,[1] and Valentine Wakelam[2]

[1] Univ. Bordeaux, CNRS, Bordeaux INP, ISM, UMR 5255, F-33400 Talence, France

[2] Univ. Bordeaux, CNRS, LAB, UMR 5804, F-33600 Pessac, France

**H-atom yield measurements**

The results of four individual runs have been coadded to obtain the curves displayed here in the upper panel of Figure S1.

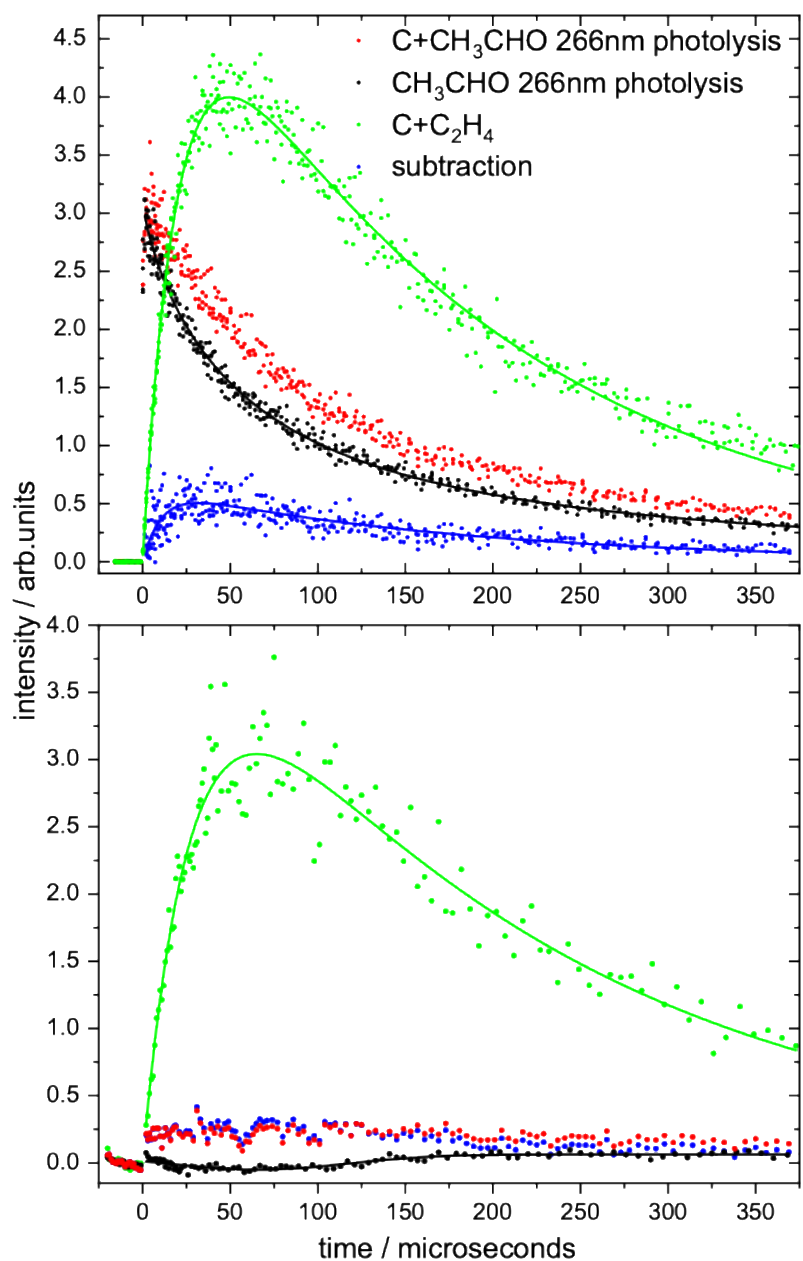

**Figure S1** H-atom fluorescence profiles as a function of reaction time recorded at 296 K. Upper panel – experiments performed with 266 nm photolysis; (red circles) H-atom signal recorded with both $CH_3CHO$ and C-atoms present in the flow; (black circles) H-atom signal recorded with $CH_3CHO$ alone in the flow; (blue circles) the resulting H-atom signal derived by subtracting the fit to the black data points from the red data points, representing the C + $CH_3CHO$ reactive signal alone; (green circles) H-atom signal from the C + $C_2H_4$ reference reaction. Lower panel - experiments performed with 212 nm photolysis. The data sets are the same as those described in the upper panel. Solid lines represent non-linear fits to the underlying data sets.

When only $CH_3CHO$ was present in the reactor (the $CBr_4$ flow was turned off), the black data points in the top panel of Figure S1 were obtained (hereafter called 'photolysis' signal), clearly showing the large H-atom signal from $CH_3CHO$ photolysis at 266 nm. When the $CBr_4$ flow was turned on, the red data points in the top panel were obtained showing that the C + $CH_3CHO$ reaction produces some additional H-atoms (hereafter called 'photolysis+reactive' signal) on top of the photolysis signal. The blue data points shown in the top panel are the result of subtracting a fit to the photolysis signal (using an arbitrary fitting function) from the photolysis+reactive signal (hereafter called 'reactive' signal), while the green data are the H-atom signal obtained from the C + $C_2H_4$ reference reaction. Both the green and the blue data are well described by expression E3. If we divide the peak amplitude of the fit to the reactive signal by the peak amplitude of the reference signal fit, and correct for the fact that the absolute H-atom yield of the reference reaction is 0.92, [46] we obtain a value for the H-atom yield of the C + $CH_3CHO$ reaction of 11.6 % at 296 K.

Considering that this result arises from a subtraction between two large signals recorded sequentially, it was necessary to consider the possibility that the recorded signal levels might have varied over the course of the experiments, leading to inaccurate results. In order to check for such a possibility, we performed additional experiments using a photolysis wavelength of 212 nm where the $CH_3CHO$ absorption cross-section is much lower with a value of $4.9 \times 10^{-22}$ $cm^2$. [39] The results of one of these experiments is shown in the lower panel of Figure S1. Here, the colour coding is the same as for the upper panel. In this case, the contribution from CH$_3$CHO photolysis is close to being negligible at this wavelength. In addition to the minor photolysis signal, the fit to the black photolysis data points here contains a small electronic noise contribution from the photolysis laser which deforms the baseline slightly at these low signal levels. When we subtract the photolysis signal from the photolysis+reactive one, we obtain a reactive signal curve which is quite different from the one derived in the upper panel as it appears instantaneously after photolysis before following a similar biexponential profile. This result is a clear indication of the instantaneous formation of atomic hydrogen by the C($^1$D) + CH$_3$CHO reaction. According to previous studies of the quenching rate of C($^1$D) atoms by N$_2$,[S1] these atoms are expected be removed rapidly in our experiments, with a calculated half-life time for C($^1$D) removal of just 0.8 μs at 296 K, so H-atom production by this reaction only contributes during the first few microseconds following photolysis. These atoms will then decay exponentially as a function of time according to expression (E1) where the decay constant $k_{1st}$ here represents mostly diffusional losses. Consequently, we can subtract this contribution from the blue curve in Figure S1 by assuming a pseudo-first-order rate constant for H-atom diffusional loss. These values can be readily obtained from the first term of the biexponential fits to the reference reaction H-atom yield data, $k_{L(H)}$ (this also corresponds to fitting to the long-time portion > 100 μs of the green curves in Figure S1), with typical values in the range 4500-5500 s$^{-1}$. If we subtract this contribution from the blue data points in Figure S1, we obtain the orange curve shown in Figure S2 alongside the uncorrected data and those of the C + C$_2$H$_4$ reference reaction.

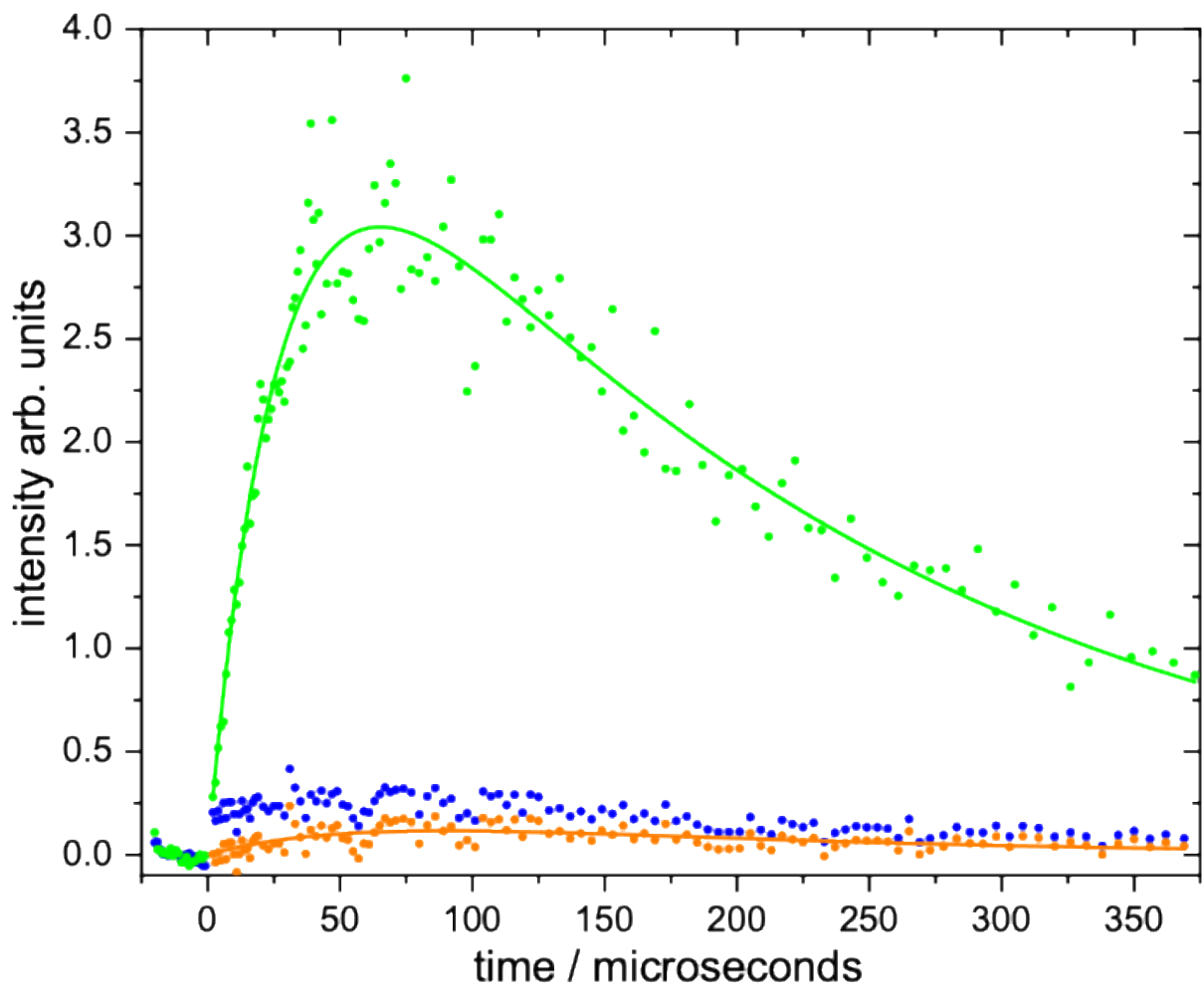

**Figure S2** H-atom fluorescence profiles as a function of reaction time recorded at 296 K for experiments performed with 212 nm photolysis. (Green circles) The C + $C_2H_4$ reference reaction; (blue circles) subtraction corresponding to the reactive signal from the C + $CH_3CHO$ reaction alone; (orange circles) the same reactive signal following subtraction of the H-atom contribution from the $C(^1D)$ + $CH_3CHO$ reaction. The solid lines represent biexponential fits to the corresponding data sets using expression (E3).

The calculated H-atom yield from this experiment was 3.5 % as described in the main text, more than three times lower than the value obtained by the 266 nm photolysis experiments. Despite the lower signal to noise ratio of the experiments conducted at 212 nm, the much lower H-atom photolysis signal of these experiments suggests that the derived H-atom yields are likely to be more reliable than those obtained from the 266 nm experiments where the subtraction of the large photolysis signal from the slightly larger photolysis+reactive one was used to yield the reactive signal.

When we use the same type of analysis for experiments performed at 177 K, to obtain the equivalent photolysis, photolysis+reactive and reference reaction data, curves of the type shown in Figure S3 are obtained.

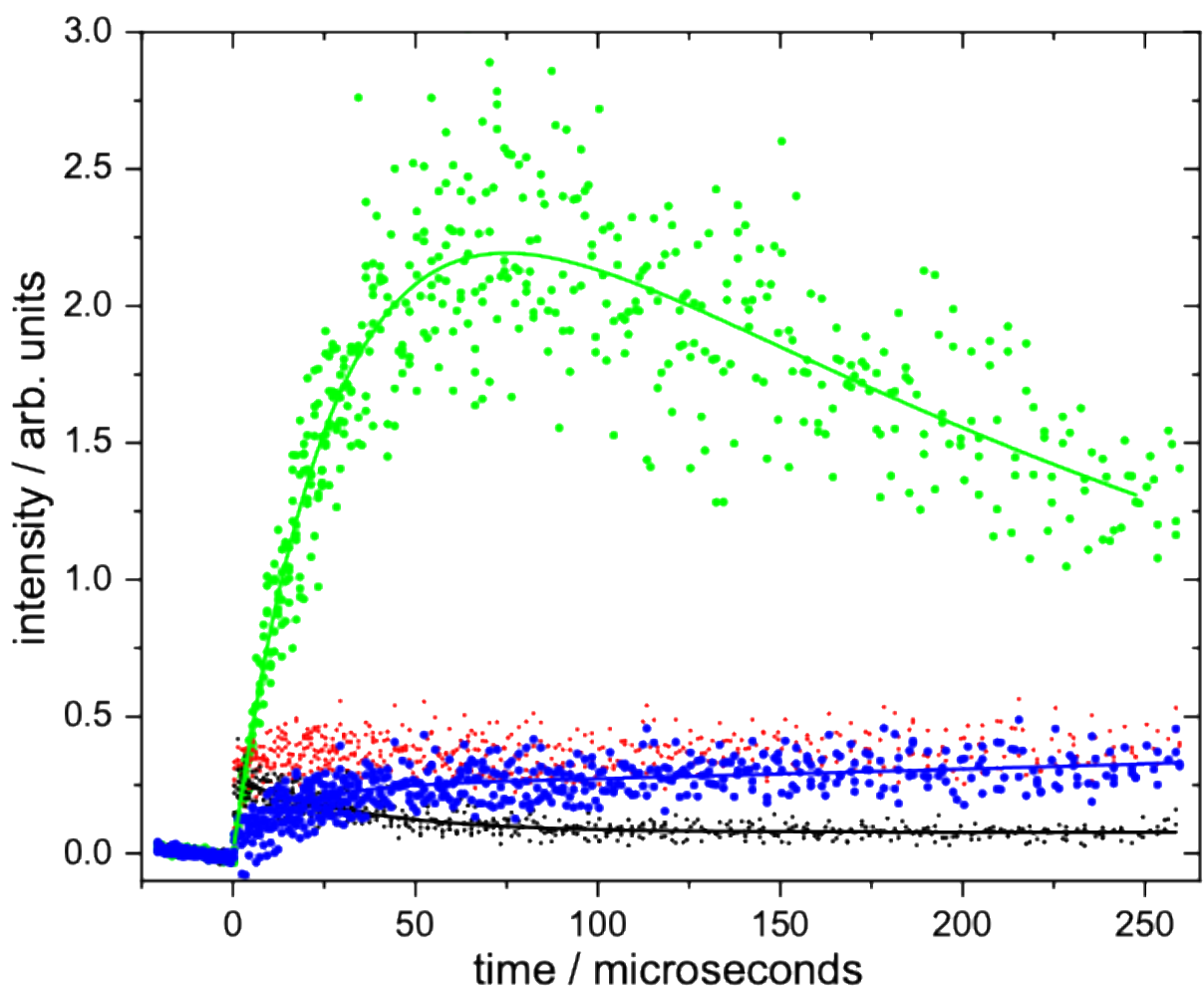

**Figure S3** H-atom fluorescence profiles as a function of reaction time recorded at 177 K for experiments performed with 212 nm photolysis. (red circles) H-atom signal recorded with both $CH_3CHO$ and C-atoms present in the flow; (black circles) H-atom signal recorded with $CH_3CHO$ alone in the flow; (blue circles) the resulting H-atom signal derived by subtracting the fit to the black data points from the red data points, representing the C + $CH_3CHO$ reactive signal alone; (green circles) H-atom signal from the C + $C_2H_4$ reference reaction.

Here it can be seen that the subtraction of the photolysis signal from the photolysis+reactive one to yield the blue reactive datapoints leads to a curve with little or no instantaneous H-atom production, so no subtraction of this contribution was applied in this case. This result is not unexpected, as $N_2$ is the carrier gas for this nozzle and the rate constant for $C(^1D)$ quenching by $N_2$ increases as the temperature decreases.[S1] However, the H-atom formation curve from the C + $CH_3CHO$ reaction represented by the blue data points in Figure S3 does not display the expected behaviour as the H-atom VUV LIF signal continues to increase with time. This behaviour is discussed in the main text.

**Frequencies (in cm$^{-1}$) of the stationary points involved in the C($^3$P) + CH$_3$CHO reaction obtained at the M06-2X/aug-cc-pVTZ level of theory.**

**Table S1 Reagents and products**

| CH$_3$CHO | C$_2$H$_4$ | CH$_3$CH | CH$_2$CHCO | CH$_2$CH | CH$_3$CCO | CH$_2$CCHO | CHCHCHO | CH$_3$C | HCO | CO |
|---|---|---|---|---|---|---|---|---|---|---|
| 125 | 340 | 153 | 142 | 717 | 34 | 183 | 145 | 505 | 1103 | 2271 |
| 511 | 340 | 771 | 328 | 873 | 168 | 217 | 285 | 864 | 1991 | |
| 754 | 714 | 996 | 549 | 943 | 503 | 412 | 651 | 1112 | 2727 | |
| 900 | 936 | 1079 | 648 | 1058 | 530 | 645 | 670 | 1305 | | |
| 1135 | 936 | 1128 | 896 | 1396 | 789 | 915 | 910 | 1313 | | |
| 1137 | 1154 | 1396 | 1017 | 1675 | 848 | 919 | 925 | 1448 | | |
| 1380 | 1437 | 1463 | 1038 | 3104 | 1064 | 949 | 946 | 2920 | | |
| 1429 | 1464 | 1467 | 1113 | 3191 | 1391 | 963 | 1037 | 3030 | | |
| 1465 | 3134 | 3010 | 1298 | 3263 | 1467 | 1391 | 1257 | 3075 | | |
| 1472 | 3138 | 3059 | 1437 | | 1473 | 1414 | 1426 | | | |
| 1863 | 3218 | 3106 | 1709 | | 1552 | 1549 | 1655 | | | |
| 2945 | 3218 | 3267 | 1959 | | 2214 | 1902 | 1834 | | | |
| 3066 | | | 3168 | | 3059 | 3026 | 2972 | | | |
| 3134 | | | 3184 | | 3105 | 3120 | 3173 | | | |
| 3188 | | | 3272 | | 3139 | 3193 | 3266 | | | |

**Table S2 Intermediates**

| C..CH$_3$CHO(1) | C..CH$_3$CHO(2) | CH$_3$CHOC-1 | CH$_3$CHOC-2 | c-CH$_3$CHOC | CHCH$_2$CHO | CH$_3$CCHO | CH$_2$CHCHO | CH$_3$CHCO |
|---|---|---|---|---|---|---|---|---|
| 81 | 36 | 181 | 109 | 157 | 106 | 114 | 170 | 96 |
| 94 | 39 | 230 | 133 | 375 | 258 | 191 | 290 | 229 |
| 151 | 151 | 270 | 322 | 440 | 270 | 233 | 376 | 305 |
| 210 | 154 | 316 | 399 | 685 | 634 | 534 | 426 | 534 |
| 511 | 521 | 583 | 522 | 889 | 668 | 884 | 571 | 692 |
| 759 | 785 | 829 | 831 | 972 | 876 | 920 | 737 | 873 |
| 899 | 902 | 993 | 1001 | 1060 | 913 | 1001 | 931 | 985 |
| 1121 | 1141 | 1087 | 1074 | 1123 | 1003 | 1087 | 1015 | 1121 |
| 1133 | 1146 | 1145 | 1147 | 1180 | 1129 | 1321 | 1075 | 1167 |
| 1376 | 1368 | 1248 | 1344 | 1334 | 1200 | 1394 | 1190 | 1359 |
| 1429 | 1426 | 1398 | 1385 | 1401 | 1319 | 1426 | 1322 | 1406 |
| 1449 | 1464 | 1450 | 1429 | 1454 | 1407 | 1448 | 1434 | 1473 |
| 1473 | 1466 | 1458 | 1467 | 1478 | 1431 | 1468 | 1451 | 1487 |
| 1864 | 1867 | 1479 | 1486 | 1500 | 1857 | 1618 | 1637 | 1847 |
| 2952 | 2946 | 3040 | 3030 | 3070 | 2950 | 2919 | 2991 | 3047 |
| 3015 | 3047 | 3085 | 3106 | 3141 | 3006 | 3031 | 3151 | 3097 |
| 3094 | 3122 | 3170 | 3164 | 3156 | 3024 | 3104 | 3155 | 3152 |
| 3178 | 3141 | 3287 | 3233 | 3178 | 3279 | 3112 | 3249 | 3169 |

**Table S3 Transition states**

| TS1 | TS2 | TS3 | TS4 | TS5 | TS6 | TS7 | TS8 | TS9 | TS10 | TS11 | TS12 | TS13 | TS14 | TS15 | TS16 |
|---|---|---|---|---|---|---|---|---|---|---|---|---|---|---|---|
| -831 | -764 | -462 | -941 | -1990 | -179 | -1928 | -564 | -1589 | -2065 | -1191 | -857 | -1700 | -717 | -549 | -865 |
| 88 | 102 | 106 | 151 | 413 | 104 | 168 | 124 | 216 | 146 | 20 | 148 | 105 | 109 | 81 | 158 |
| 112 | 169 | 188 | 159 | 746 | 230 | 219 | 194 | 244 | 333 | 341 | 302 | 278 | 228 | 237 | 288 |
| 237 | 281 | 351 | 419 | 778 | 352 | 263 | 293 | 310 | 440 | 465 | 325 | 323 | 237 | 288 | 345 |
| 343 | 363 | 367 | 431 | 994 | 518 | 550 | 509 | 502 | 625 | 881 | 483 | 555 | 523 | 678 | 386 |
| 500 | 563 | 862 | 908 | 1099 | 841 | 697 | 691 | 911 | 645 | 914 | 496 | 684 | 537 | 693 | 570 |
| 942 | 958 | 986 | 963 | 1243 | 892 | 848 | 708 | 974 | 845 | 1013 | 628 | 873 | 594 | 742 | 743 |
| 999 | 989 | 1102 | 1007 | 1411 | 1033 | 928 | 855 | 1054 | 968 | 1110 | 880 | 989 | 818 | 915 | 783 |
| 1130 | 1123 | 1123 | 1127 | 2260 | 1064 | 992 | 897 | 1110 | 1021 | 1218 | 910 | 1089 | 996 | 935 | 911 |
| 1202 | 1180 | 1349 | 1222 | 3120 | 1283 | 1076 | 1032 | 1379 | 1104 | 1318 | 955 | 1160 | 1039 | 1112 | 1038 |
| 1397 | 1397 | 1400 | 1374 | 3198 | 1404 | 1366 | 1113 | 1410 | 1126 | 1384 | 968 | 1187 | 1392 | 1165 | 1120 |
| 1460 | 1461 | 1430 | 1430 | 3224 | 1419 | 1407 | 1382 | 1449 | 1297 | 1447 | 1390 | 1322 | 1452 | 1380 | 1205 |
| 1471 | 1471 | 1466 | 1462 | | 1579 | 1453 | 1543 | 1486 | 1410 | 1476 | 1410 | 1425 | 1464 | 1398 | 1410 |
| 1539 | 1542 | 1480 | 1478 | | 1747 | 1662 | 1892 | 1603 | 1751 | 1505 | 1567 | 1740 | 1509 | 1847 | 1616 |
| 3029 | 3019 | 3044 | 3037 | | 2218 | 2275 | 2863 | 1993 | 1849 | 3045 | 1807 | 2235 | 2062 | 2346 | 1821 |
| 3098 | 3088 | 3101 | 3092 | | 3026 | 2929 | 3112 | 3011 | 3102 | 3119 | 3019 | 2965 | 3039 | 2981 | 2984 |
| 3133 | 3101 | 3167 | 3173 | | 3107 | 3145 | 3204 | 3121 | 3185 | 3142 | 3122 | 3198 | 3098 | 3047 | 3132 |
| 3227 | 3284 | 3206 | 3233 | | 3204 | 3266 | 3264 | 3153 | 3203 | 3175 | 3199 | 3264 | 3116 | 3125 | 3270 |

**Cartesian coordinates of the reagents, products, intermediates and transition states involved in the reaction between C($^3$P) and CH$_3$CHO.**

**Table S4 Reagents and products**
$^1$CH$_3$CHO
```
  C   0.00024857575090    0.46153310210987    0.00000102337167
  C  -0.92856863176243   -0.71492544390163    0.00000018366751
  O   1.19742506313383    0.38033986206513   -0.00000156302675
  H  -0.48755573680517    1.45495315729974    0.00000054024278
  H  -0.36647653479386   -1.64469747136120    0.00000056328423
  H  -1.57641162791540   -0.66412554404021    0.87700796189951
  H  -1.57641010760786   -0.66412566217169   -0.87700870943894
```

$^3$C$_2$H$_4$
```
  C   0.00000006537773    0.00000033277720    0.72338341984640
  C   0.00000005431114    0.00000155533885   -0.72338333865270
  H   0.00000040903801    0.92220565048982    1.29160720328716
  H  -0.00000054237644   -0.92220576770063    1.29160584539679
  H   0.92220556199469   -0.00000040974514   -1.29160666344576
  H  -0.92220554834514   -0.00000136116010   -1.29160646643189
```

$^3$CH$_3$CH
```
  C  -1.65379952969810   -0.06565414552685    0.08919789616004
  C  -1.32802765343951   -0.88520308187686    1.25796167539399
  H  -1.31072554063688    0.96092358195386    0.22981288909705
  H  -1.17721911637490   -0.44496667062508   -0.82139569268961
  H  -2.73263201781374   -0.02349556352683   -0.09757359448953
  H  -1.43964218743688   -1.93124601449822    1.49081463102807
```

$^2$CH$_2$CHCO
```
  C   0.14738261146356    1.21829691749676    1.10805479570022
  C  -1.01408046787991   -0.60968400531512    0.00977986937116
  O   0.50638647752991    2.32662385626529    1.27166007121992
  H  -0.63577694787311   -1.27246961630269    0.77742015260660
  H  -1.63973479068776   -1.03233513041897   -0.76417135987455
  C  -0.70983057714482    0.68054098756440    0.03030921594027
  H  -1.06410509550787    1.38450592741032   -0.71611965246362
```

$^2$CH$_2$CH
```
  C  -0.50135275862908   -0.04474122246490    1.33843614695690
  C  -1.62055631794144    0.11576283099420    0.69395868452030
  H  -0.13475593835429    0.13114456406783    2.33669971071704
  H  -2.50464659944958    0.53084685199416    1.17520942950268
  H  -1.72122867262561   -0.15654492199129   -0.35134416229693
```

$^2$CH$_3$CCO

|   |   |   |   |
|---|---|---|---|
| C | 0.10042726297591 | 1.16860803964083 | 0.84959068127642 |
| C | -0.08428685218921 | 0.65084016145083 | 1.97652561195830 |
| C | 0.29607935458759 | 1.67443902585970 | -0.49495069410729 |
| H | 0.47647160426826 | 2.74781314472088 | -0.52451028986164 |
| H | 1.13483838574479 | 1.16038571642066 | -0.96952583834220 |
| H | -0.58335795095514 | 1.44900852239294 | -1.10306285507223 |
| O | -0.25638780443220 | 0.18201638951415 | 3.04976738414865 |

$^2$CH$_2$CCHO

|   |   |   |   |
|---|---|---|---|
| C | 0.96154308118092 | 1.12168439393746 | 0.76487244265164 |
| C | -0.45437696362573 | -1.03799077141873 | 0.02361333175363 |
| O | 0.45891054304112 | 2.23056225595666 | 0.94566598043344 |
| H | 2.03955653691590 | 0.95703103164151 | 0.89744212786174 |
| H | -0.89368465633321 | -1.70970251647054 | 0.75568869885098 |
| C | 0.22176476322261 | -0.00097679133257 | 0.38416662142123 |
| H | -0.61560843770160 | -1.27018623101379 | -1.02534322447267 |

$^2$CHCHCHO

|   |   |   |   |
|---|---|---|---|
| C | 0.02505138487338 | 0.38023363052660 | -0.07983286808672 |
| C | -0.61601346285825 | -0.87944147672973 | -0.57026347455781 |
| O | 1.14295120514227 | 0.44987031535725 | 0.35176784527285 |
| H | -0.61316417435911 | 1.27930592358194 | -0.13650219847253 |
| H | -1.62378729746943 | -0.79300489780833 | -0.96365470261471 |
| C | 0.00291666086090 | -2.03117002747002 | -0.52523090160643 |
| H | 0.96865203731024 | -2.37073054235771 | -0.18627131403465 |

$^2$CH$_3$C

|   |   |   |   |
|---|---|---|---|
| C | -0.84276275481982 | -1.10767295030575 | -0.40760661809945 |
| H | -1.50733888174665 | -0.67489628634247 | 0.34748545812368 |
| H | -1.45338097659153 | -1.82602484740529 | -1.00090414991871 |
| C | -0.51484569627681 | -0.28302361935983 | -1.53849044368389 |
| H | -0.02574931486519 | -1.68225954828665 | 0.04125351517837 |

$^2$HCO

|   |   |   |   |
|---|---|---|---|
| C | -0.75929793578901 | 0.46706476686173 | 3.74744085413021 |
| O | -0.88703468603407 | -0.60910982590768 | 3.31159170544752 |
| H | -0.22644737817694 | 1.30021505904595 | 3.22147744042225 |

$^1$CO

|   |   |   |   |
|---|---|---|---|
| C | 0.00000000000000 | 0.00000000000000 | -0.64095512334366 |
| O | 0.00000000000000 | 0.00000000000000 | 0.48043012334366 |

**Table S5 Intermediates**
C..CH$_3$CHO(1)

|   |   |   |   |
|---|---|---|---|
| C | 0.37827096379003 | 1.21425758081006 | -0.05719722015459 |
| C | -0.15521059607553 | -0.19142598601603 | -0.06034724698104 |
| O | 1.54273638830977 | 1.49530952826819 | -0.06839837954573 |
| H | -0.38989370195627 | 2.01035577970937 | -0.04387683453547 |
| H | -0.76796125793649 | -0.34052703964795 | 0.83229481909136 |
| H | -0.80525394277769 | -0.33434545013003 | -0.92523685215610 |
| C | -0.59652941321533 | -2.84460708990745 | 0.40198700545542 |
| H | 0.67515856447244 | -0.89812043274761 | -0.08559309389103 |

C..CH$_3$CHO(2)

| | | | |
|---|---|---|---|
| C | 0.64077877937034 | 0.71368735458205 | -0.04694913246017 |
| C | -0.13979816553787 | -0.49547721681412 | -0.48229614420737 |
| O | 0.41850861615930 | 1.82587499907692 | -0.43576239764411 |
| H | 1.47218844260568 | 0.52015782895997 | 0.65651700834465 |
| H | -0.33020618527620 | -1.17098888566534 | 0.36112670371252 |
| H | -1.06660514233524 | -0.20156975847138 | -0.96728580710621 |
| C | -1.34201162054058 | -0.05862557086836 | 1.93057665280537 |
| H | 0.47289144666365 | -1.07261010862286 | -1.17983754326526 |

CH$_3$CHOC-1

| | | | |
|---|---|---|---|
| C | -0.06169637062640 | 0.51109616887102 | 0.08831010514558 |
| C | -0.87611719381657 | -0.71004063722834 | 0.00638826629361 |
| O | 1.30890096031773 | 0.44701153550483 | -0.09122266576236 |
| H | -0.40245827926878 | 1.51229531450892 | 0.28067773875815 |
| H | -0.55785387149294 | -1.44787496314362 | 0.74872405214579 |
| H | -1.92134949868952 | -0.46255073592807 | 0.17092815259011 |
| H | -0.76770736718025 | -1.19210450641454 | -0.96959446909989 |
| C | 1.93757445946572 | -0.63697120723878 | -0.33298553647375 |

CH$_3$CHOC-2

| | | | |
|---|---|---|---|
| C | -0.13488465237646 | 0.58149715499543 | 0.15811433250295 |
| C | -0.85163449365496 | -0.68749217858913 | -0.04996525500728 |
| O | 1.24021795464936 | 0.54230867547619 | -0.06474552754905 |
| H | -0.55563247040100 | 1.57487287098870 | 0.19495976904104 |
| H | -0.34020375524561 | -1.50636547913879 | 0.45730928208388 |
| H | -1.86411388761913 | -0.60992643341157 | 0.33925249343448 |
| H | -0.91563183464711 | -0.94806320745382 | -1.11221256799774 |
| C | 1.95381417932909 | 1.59276088648779 | -0.11120285440027 |

c-CH$_3$CHOC

| | | | |
|---|---|---|---|
| C | -1.42669916821314 | 0.07852219263223 | -0.17251357240593 |
| C | -0.10382926450088 | -0.24805702189098 | 0.43257258015879 |
| H | -1.88674934362760 | 0.91860877468082 | 0.34872222593876 |
| H | -1.30662798303085 | 0.32117000047945 | -1.22550781735742 |
| H | -2.09061311374102 | -0.78163360637335 | -0.07873341847405 |
| H | -0.08241524038929 | -0.42020509745048 | 1.50390118887227 |
| C | 1.03973020896518 | -0.66752244848650 | -0.34582475430118 |
| O | 1.03630890453760 | 0.62349320640881 | -0.00424943243124 |

CHCH$_2$CHO

| | | | |
|---|---|---|---|
| C | -0.01602777734165 | 0.50611272670787 | -0.00016808114833 |
| C | -0.93037592275407 | -0.70319421068969 | -0.00231685920684 |
| O | 1.17996531962586 | 0.46072479305089 | -0.00431769815120 |
| H | -0.54787760168306 | 1.47608855817166 | 0.00539524516904 |
| H | -1.59173362776372 | -0.59186372757565 | 0.86820020776613 |
| H | -1.59537252203767 | -0.58500817666733 | -0.86938869703005 |
| C | -0.23350814339171 | -1.98789210594971 | -0.00739398779598 |
| H | 0.80915790074603 | -2.25355838194804 | -0.02638722290278 |

CH$_3$CCHO

| | | | |
|---|---|---|---|
| C | 0.67980366909451 | 1.23205295240113 | 0.45997766074456 |
| C | -0.62602354217283 | -0.93958074561071 | -0.30410501394364 |
| O | 0.64875751826923 | 2.41204828954016 | 0.77497712905572 |
| H | 1.64276990674396 | 0.68302586377604 | 0.43725890759942 |
| H | -1.28041864951173 | -1.47026230570059 | 0.39105849708508 |
| H | -1.08436274121946 | -1.00317030192436 | -1.29340715508028 |
| C | -0.44562666966826 | 0.45198455441610 | 0.10233650411110 |
| H | 0.33535737544925 | -1.46214318014993 | -0.33660415425232 |

CH$_2$CHCHO

| | | | |
|---|---|---|---|
| C | 0.68154353300046 | 1.24556316638135 | 0.60174082501547 |
| C | -0.49491197335405 | -0.91396229828975 | -0.03696662762339 |
| O | 0.69459562412003 | 2.45934377005440 | 0.74052720741157 |
| H | 1.57197609333649 | 0.63966958499855 | 0.83758331741769 |
| H | -0.84894708273182 | -1.56438189995159 | 0.75139124954793 |
| C | -0.46951686342105 | 0.52520318617813 | 0.13653225051874 |
| H | -0.27556830773846 | -1.36246964599986 | -0.99646929293493 |
| H | -1.34265073661161 | 1.13064412642877 | -0.09214012165306 |

CH$_3$CHCO

| | | | |
|---|---|---|---|
| C | 0.60080389073852 | 1.19563131206634 | 0.46832727965372 |
| C | -0.64068672076718 | -0.90754769278441 | -0.28521753027353 |
| O | 0.75045196649659 | 2.32462148097454 | 0.80760136627814 |
| H | -1.25106703476208 | -1.49985537471924 | 0.40220357120265 |
| H | -1.10064197160255 | -1.01510101403683 | -1.27105209419195 |
| C | -0.61569074736025 | 0.51062333341171 | 0.12712390464943 |
| H | 0.36364358022857 | -1.32337142737512 | -0.31370679144967 |
| H | -1.53912296297162 | 1.08050938246300 | 0.17746029413119 |

**Table S6 Transition states**

TS1

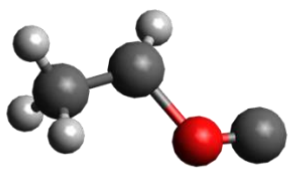

```
C  -0.05720564279610    0.29538117831612    0.17285418477201
C  -1.08355844047487   -0.73401738061609   -0.00395695052990
O   1.49355261810905   -0.29993479450514   -0.01025864542276
H  -0.06248892240336    1.35463643127164   -0.03739049937801
H  -0.70412250889868   -1.69492548768886    0.34539216111954
H  -1.98272075070084   -0.49392255678816    0.56678387918892
H  -1.37467509971416   -0.84790509171746   -1.05362750817948
C   2.43051174687896    0.44155070172796   -0.07857162157032
```

TS2

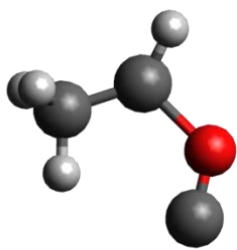

```
C  -0.15223961884739    0.53567330572562    0.15072759474295
C  -0.91460486297296   -0.70703228220588    0.05647005700796
O   1.42036978400442    0.45324651428865   -0.31585308052880
H  -0.24380034920169    1.37701293406326    0.81441404241461
H  -1.14736336271257   -1.11769848563447    1.04228804182769
H  -1.85553240905283   -0.56460435568992   -0.47926367344120
H  -0.32138787344784   -1.44527405285381   -0.49464505510154
C   1.87385153093986   -0.51046260876202   -0.87291228332440
```

TS3

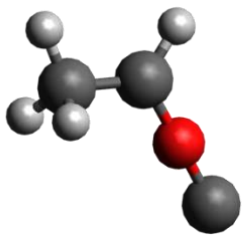

```
C   0.12944085330201    0.63588279529890    0.07187846419314
C  -0.83393178067313   -0.48430300510953    0.04125716338321
O   1.41225369305669    0.38244928516441   -0.13287442998109
```

| | | | |
|---|---|---|---|
| H | -0.11508825226514 | 1.66445244954202 | 0.29736229513426 |
| H | -0.62714433576298 | -1.20167822160745 | 0.84074931411581 |
| H | -1.84611601710543 | -0.10680966527831 | 0.16455903446725 |
| H | -0.76990111083459 | -1.02743763323224 | -0.90451734546930 |
| C | 2.65048695028257 | 0.13744399522220 | -0.37841449584328 |

TS4

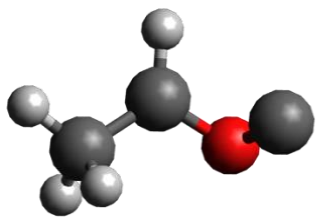

| | | | |
|---|---|---|---|
| C | -1.48480241025165 | 0.16244175273011 | -0.14306477848817 |
| C | -0.14419689907320 | 0.15865750220034 | 0.45663515692257 |
| H | -2.25430484603316 | 0.29422903893720 | 0.61242513859876 |
| H | -1.56455505753627 | 0.95888355918588 | -0.88771622348291 |
| H | -1.65009156119415 | -0.78545209869439 | -0.66514833259692 |
| H | 0.08306532531388 | 0.23224419959877 | 1.50902797613579 |
| C | 1.26371213990052 | -1.21958912877773 | -0.02690822871486 |
| O | 0.93027830887403 | 0.02296117481984 | -0.39688370837425 |

TS5

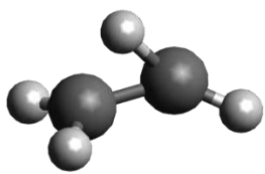

| | | | |
|---|---|---|---|
| C | 0.17310781936947 | 0.42071935133850 | -0.34641715259160 |
| C | -0.16990589134170 | -0.47718870048394 | 0.72951050746345 |
| H | 0.42696947932718 | 1.45523018071552 | -0.14556657074040 |
| H | 0.43937364007232 | 0.01328540340769 | -1.31522104887663 |
| H | -1.05033375705522 | 0.00447371713848 | -0.05483197789659 |
| H | 0.18078870962794 | -1.41651995211624 | 1.13252624264178 |

TS6

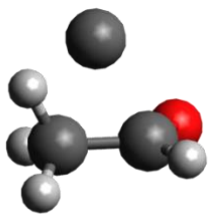
48

| | | | |
|---|---|---|---|
| C | 0.56031360993020 | 0.75074566596303 | 0.04812001815998 |
| C | -0.19771242379959 | -0.47594477324034 | -0.41756592756580 |
| O | 0.23847180064310 | 1.86074572172681 | -0.29910991974309 |
| H | 1.47727703905022 | 0.56114025963971 | 0.62260130690073 |
| H | -0.95638468413506 | -0.94391872293141 | 0.34279039330416 |
| H | -0.86267520569687 | -0.20688042903043 | -1.23498061704550 |
| C | -0.71478369258592 | -0.22769764182752 | 1.57115931631907 |
| H | 0.45549355659394 | -1.31819008029985 | -0.63301457032956 |

TS7

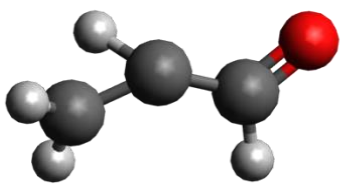

| | | | |
|---|---|---|---|
| C | 0.64459142196413 | 1.21621735640618 | 0.36398936585795 |
| C | -0.49767242455077 | -1.15103574512559 | -0.12664852412790 |
| O | 0.49589858422241 | 2.42556334417655 | 0.37826458796669 |
| H | 1.63799241758782 | 0.76424852136766 | 0.55216987295967 |
| H | -1.38130128368981 | -1.67694423096363 | 0.20413240765649 |
| C | -0.38336772384073 | 0.26006361287014 | 0.10261927111954 |
| H | 0.35115380616565 | -1.71116933548506 | -0.50095754132524 |
| H | -0.86729479785870 | -0.12694352324626 | -0.97356944010720 |

TS8

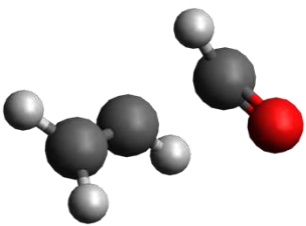

| | | | |
|---|---|---|---|
| C | 0.76523866294568 | 1.34008588386584 | 0.72386697251320 |
| C | -0.54372434151370 | -0.80078864976117 | -0.05718581389702 |
| O | 1.16574144185388 | 1.97773503728122 | -0.17884335013954 |
| H | 1.36307242504155 | 0.68783977154003 | 1.39372237424388 |
| H | -0.38133890655763 | -1.69335466449329 | 0.53656418419221 |
| C | -1.06539922763722 | 0.29952123231536 | 0.47362076996040 |
| H | -0.19038548406763 | -0.82539058287546 | -1.08594706188425 |
| H | -1.59668428346494 | 1.17396196192746 | 0.13640073271114 |

TS9

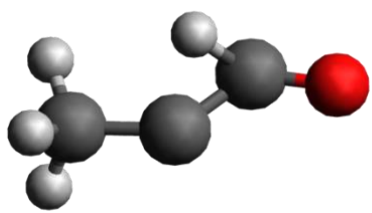

```
C   0.73918502334895    1.28578168064574    0.47929704170810
C  -0.33457674668249   -1.01170539452465   -0.22052307506800
O   0.74867714034847    2.32922005701269    1.11302805201257
H  -0.27281687442774   -1.80637311623683    0.52146225262229
H  -1.27929369702892   -1.11529331627524   -0.75141834524252
C  -0.21599915449573    0.29777233482378    0.44215246708388
H   0.48812122669656   -1.13198240972713   -0.93729685811244
H   0.12670308224090    1.15258016428164   -0.64670153500388
```

TS10

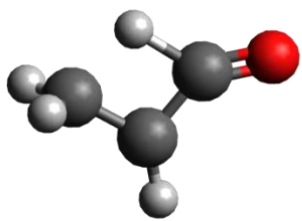

```
C   0.87242263558529    0.82653726981531    0.33461642967351
C  -0.36878256389876   -0.90421863618385   -0.30052424822514
O   1.54138346818488    1.75050288178775    0.67205755001027
H  -0.70717714882187   -1.66444232141961    0.39958120616345
H  -0.53377946143686   -1.15948707290775   -1.34469623276925
C  -0.50470164728930    0.51276211736106    0.09615809708273
H   1.08135423219492   -0.50532526609047   -0.03290098957740
H  -1.38071951451831    1.14367102763754    0.17570818764183
```

TS11

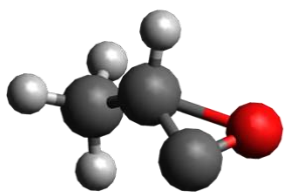

```
C  -0.86851886169218    0.08514289952413   -0.24283058921878
C   0.44793863700300   -0.28211645920196    0.33567896530872
H  -1.30981383213753    0.92278416984308    0.29556646994893
H  -0.78629206494699    0.31861406096771   -1.29992495456323
H  -1.53924852748814   -0.77107837310868   -0.12002373561791
```

H   0.51970616504294    -0.33917113115782    1.41777174679272
C   1.61990158910297    -0.55844126873624   -0.41159378279536
O   1.91632689511594     0.62426610186978    0.02535588014492

TS12

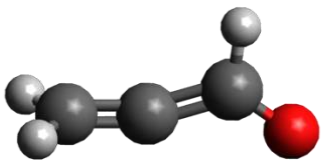

C   0.77707373201238     1.24240793679374    0.63090148104083
C  -0.49499309585804    -1.01414244319077   -0.04797557615437
O   0.47052892053717     2.42964431359929    0.63989632421312
H   1.77969007597565     0.91626470856360    0.94294305697101
H  -1.03012921819394    -1.61553901468695    0.68060421205746
C  -0.06135215524271     0.17867640666956    0.24662336977080
H  -0.41917769697762    -1.40563660171930   -1.05806594865580
H  -1.50512027565291     1.42793468377082   -0.09272811154306

TS13

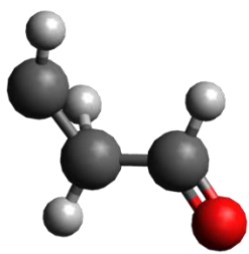

C   0.39799785166907     1.03908380398576   -0.13206806456664
C  -0.10839859162136    -1.43101586031473    0.30603997229754
O   0.37284872396081     2.01104680132472   -0.85853252918593
H   0.96637197540833     1.04764653968492    0.81498579110423
H  -1.02026532167987    -0.49672768782768    0.57344137710394
C  -0.30786468066759    -0.19647235447199   -0.42810473238656
H   0.58838624962613    -1.74353128544745    1.06584744589918
H  -0.88907620669553    -0.23002995693355   -1.34160926026576

TS14

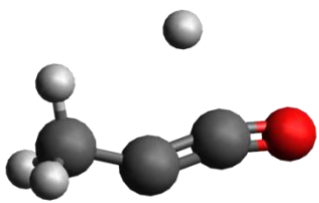

```
C    0.28372106185905     1.45179559316750     0.45973095477091
C   -0.57269641814430    -0.92759522666837    -0.31161401254181
O    0.68374699554479     2.50223564998397     0.80615787421537
H    1.82825719129217     0.29775426458240     0.42850905785293
H   -1.19179734205543    -1.45119910486468     0.42113421625755
H   -1.09341724218976    -1.01802485480927    -1.26611582687102
C   -0.46334780850551     0.47446793433977     0.06842832568178
H    0.39579042918369    -1.42547912898347    -0.37473821404608
```

TS15

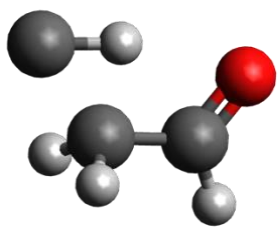

```
C    0.38702651850916     0.97309770629321    -0.02237630265117
C   -0.46994248672815    -0.26808505457268    -0.01948735885215
O    1.58481511601571     0.94773086942132    -0.00466847143509
H   -0.16897287856313     1.92711781526219    -0.05147667222412
H   -1.15666602120808    -0.24093964782953     0.83077516266035
H   -1.01388875213268    -0.34284122423404    -0.96288530533835
C    0.03786813859075    -1.92806373248779     0.25701095677716
H    0.79976036551643    -1.06801673185269    -0.02689200893662
```

TS16

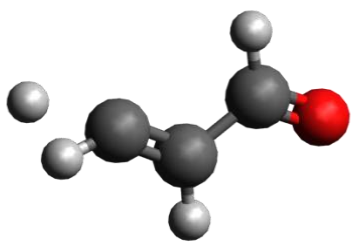

```
C    0.72029266785694     1.30983817082665     0.64841339574676
C   -0.15450638633619    -0.72766612048569    -0.36555203238590
O    0.58985044726914     2.46337581593982     0.96290694371913
H    1.65275174030098     0.75616784471709     0.85124128759293
H   -1.05303856342586    -1.75154993917116     1.07191044213191
C   -0.33671164134523     0.53483882358284    -0.03525331878361
H   -0.62662761140621    -1.48065955600143    -0.97459900908704
H   -1.27549036631358     1.05526495039185    -0.21686890123417
```